\documentclass[10pt]{article}

\usepackage{amsmath}
\usepackage{amssymb}

\usepackage{graphicx}

\usepackage{cite}

\usepackage{color} 

\usepackage[labelfont=bf,labelsep=period,justification=raggedright]{caption}

\makeatletter
\renewcommand{\@biblabel}[1]{\quad#1.}
\makeatother

\date{}

\begin{document}

\begin{flushleft}
{\Large
\textbf{High-resolution transcriptome analysis with long-read RNA sequencing}
}
\\
Hyunghoon Cho$^{1,\ddagger }$, 
Joe Davis$^{2}$, 
Xin Li$^{3}$,
Kevin S. Smith$^{3}$, 
Alexis Battle$^{1,2,\ast}$,
Stephen B. Montgomery$^{1,2,3,\ast}$
\\
\bf{1} Department of Computer Science, Stanford University, Stanford, CA, USA
\\
\bf{2} Department of Genetics, Stanford University, Stanford, CA, USA
\\
\bf{3} Department of Pathology, Stanford University, Stanford, CA, USA
\\
$\ast$ E-mail: smontgom@stanford.edu and ajbattle@cs.stanford.edu
\\
$\ddagger$ Present address: Computer Science and Artificial Intelligence Laboratory, Massachusetts Institute of Technology, Cambridge, MA, USA
\end{flushleft}

\section*{Abstract}
RNA sequencing (RNA-seq) enables characterization and quantification of individual transcriptomes as well as detection of patterns of allelic expression and alternative splicing. Current RNA-seq protocols depend on high-throughput short-read sequencing of cDNA. However, as ongoing advances are rapidly yielding increasing read lengths, a technical hurdle remains in identifying the degree to which differences in read length influence various transcriptome analyses. In this study, we generated two paired-end RNA-seq datasets of differing read lengths (2x75 bp and 2x262 bp) for lymphoblastoid cell line GM12878 and compared the effect of read length on transcriptome analyses, including read-mapping performance, gene and transcript quantification, and detection of allele-specific expression (ASE) and allele-specific alternative splicing (ASAS) patterns. Our results indicate that, while the current long-read protocol is considerably more expensive than short-read sequencing, there are important benefits that can only be achieved with longer read length, including lower mapping bias and reduced ambiguity in assigning reads to genomic elements, such as mRNA transcript. We show that these benefits ultimately lead to improved detection of \textit{cis}-acting regulatory and splicing variation effects within individuals. 


\section*{Author Summary}

Ongoing improvements to next generation sequencing technologies are leading to longer sequencing read lengths, but a thorough understanding of the impact of longer reads on RNA sequencing analyses is lacking. To provide a better understanding of this issue, we generated and compared two RNA sequencing datasets of differing read lengths (2x75 bp and 2x262 bp) and investigated the impact of read length on various aspects of analysis, including the performance of currently available read-mapping tools, gene and transcript quantification, and detection of allele-specific expression patterns. Our results indicate that, while the scalability of read-mapping tools and the cost-effectiveness of long read protocol is an issue that requires further attention, longer reads enable more accurate quantification of diverse aspects of gene expression, including individual-specific patterns of allele-specific expression and alternative splicing.

\section*{Introduction}

The application of next-generation sequencing (NGS) to RNA has provided more complete means to annotate and quantify transcriptomes. Specifically, it has improved the characterization of many aspects of RNA biology including the detection of transcription start sites \cite{Plessy2010, Ni2010, Valen2009}, allele-specific expression \cite{Montgomery2010}, alternative splicing events \cite{Sultan2008}, fusion transcripts \cite{Levin2009}, RNA-editing \cite{Peng2012}, and antisense transcription \cite{Cloonan2008}. However, the analytical properties of RNA-seq data with respect to rapidly changing technological parameters, such as continually increasing read lengths, remain challenging to ascertain. While it would be ideal to sequence the full length of each mRNA molecule, current NGS technologies are limited to analyzing short fragments of cDNA, and only a limited number of bases can be read from each fragment with reasonable accuracy. Paired-end sequencing methods were developed to partially address this limitation, and, notably, it has been shown that the paired-end reads enable more accurate estimations of transcript abundances \cite{Salzman2011}. Moreover, it is established that longer reads will improve performance in several of the noted applications of RNA-seq, such as transcript assembly, transcript quantification, and gene fusion detection \cite{Shendure2008, Ozsolak2011}. However, as public RNA-seq data of variable read lengths are increasingly available to the research community, ongoing assessment of the effect of read length on various aspects of transcriptome analysis will be necessary to understand the practical implications of integrating and comparing such data. In this study, we generated paired-end RNA-sequencing datasets for the lymphoblastoid cell line GM12878 with 262 bp reads (L262) and 75 bp reads (L75) and carried out a comparative analysis. 

One aspect of RNA-seq analysis we focused on was its ability to quantify gene expression. Although RNA-seq is considered to be a more robust quantification method than hybridization-based microarray assays due to its wide dynamic range and ability to detect novel sequences, various sources of bias inherent to RNA-seq have also been identified, including GC-content \cite{Benjamini2012}, transcript length \cite{Oshlack2009}, overdispersion \cite{Marioni2008} and mappability \cite{Schwartz2011}. While biases arising from GC-content and transcript length can be countered with appropriate normalization methods  \cite{Benjamini2012, Oshlack2009}, the inability of short reads to effectively disambiguate fragments from non-unique regions of the genome can only be mitigated by using longer reads. To evaluate the benefit of longer reads, we compared the number of reads that uniquely mapped to each gene in both L262 and L75 datasets and investigated the differences that are distinguishable from random biological variation. We found that the proportion of reads in L262 that unambiguously aligned to a pseudogene is roughly three times higher than that of L75, with a similar observation in coding genes with low mappability. 

Quantification of transcript-level expression from RNA-seq data is a considerably harder problem than that of gene-level expression, since alternative splicing patterns introduce  ambiguity in mapping each read to a particular transcript. To estimate transcript abundances from short-read data, several methods based on statistical inference have been developed such as Cufflinks \cite{Trapnell2010}, TIGAR \cite{Nariai2013}, RSEM \cite{Li2011} and FluxCapacitor (http://ux.sammeth.net/capacitor.html). While these methods aim to make an informed guess using available information, one could improve the outcome or even make these tools obsolete by using longer reads to reduce or remove the overall ambiguity in mapping reads to transcripts. In this study, we looked at the number of mRNA isoforms each read maps to and found that long reads lead to better specificity. In particular, 32.98\% of exonic reads in L262 were unambiguously assigned to a single isoform, compared to 25.04\% for L75. To further evaluate this class of methods, we evaluated the degree to which two different methods, Cufflinks and FluxCapacitor, predicted transcript abundances from the L75 data that were directly measured in the L262 data.  

RNA-sequencing also supports the quantification of allele-specific expression (ASE) and allele-specific alternative splicing (ASAS), providing direct evidence of \textit{cis}-regulatory effects within an individual.  Analysis of allele-specific patterns has the potential to reveal additional regulatory variants and mechanisms, but analysis in RNA-seq is limited to evaluation of reads containing at least one heterozygous site, which is needed to differentiate between the two copies of a gene. This significantly reduces the effective library size and consequentially many tests are statistically underpowered. Therefore, long reads are useful due to a higher chance of containing a heterozygous site.  Here, we report the degree of improved detection of ASE and ASAS from longer read data. We further identified compelling cases of ASAS, which give us a glimpse of the complex regulatory mechanisms within individuals.

One important practical consideration regarding read length is cost-efficiency.  Specifically, for any given analytical goal, it is not \textit{a priori} clear whether it is more cost-effective to use a long read protocol, or a short read protocol with higher depth. We provide some insight into this issue by carrying out a cost comparison between L262 and L75, generated on a MiSeq and HiSeq, respectively. Based on current sequencing prices, it is more cost-effective to select a short-read protocol for many analyses. Specifically, obtaining a target number of reads with isoform-specificity can be achieved at lower total cost with short reads, even though a larger number of reads must be sequenced; a similar result is seen for obtaining haplotype-specific reads for ASE analysis. On the other hand, the reduction in mapping bias, which ultimately leads to more accurate quantification, is only accessible via longer read lengths and cannot be achieved by simply increasing the number of short reads.  More broadly, some elements (genes, transcripts, or loci) in low mappability regions would remain under-represented in short read data, regardless of depth or cost.  As the price gap between long-read and short-read RNA sequencing platforms becomes smaller, we expect the benefits of longer read lengths to become more apparent.

\section*{Materials and Methods}
\subsection*{Sample preparation and sequencing}

GM12878 is a CEPH/UTAH family EBV transformed peripheral blood B lymphocyte cell line (Cat\# XC01463) purchased from The Coriell Institute (Camden, NJ) and grown in RPMI 1640 supplemented with 10\% FCS and penicillin/streptomycin in humidified 5\% CO2 at a concentration of approximately $1x10^{6}$ cells/ml. Total RNA was isolated from $1x10^{7}$ GM12878 cells using Trizol. RNA quality was assessed with the Agilent Bioanalyzer 2100 and RIN scores above 9 were used for cDNA production.  One microgram total RNA was used to isolate poly A purified mRNA which was used for library construction using the Illumina TruSeq RNA Preparation Kit. RNA fragmentation time was 8 minutes for samples to be used for 75 bp paired-end reads and 1 minute for 262 bp paired-end reads. Strand specificity was performed using dUTP during second strand synthesis.  GM12878 cDNA was indexed with Illumina adapters for use in multiplex HiSeq runs. GM12878 cDNA sequenced by an Illumina HiSeq yielded 25,933,924 75 bp paired-end reads (L75). For MiSeq sequencing, 6 pM cDNA was denatured and loaded into a 500 cycle MiSeq sequencing cartridge that yielded 9,524,186 262 bp paired-end reads (L262). All data is freely available from GEO under the accession identifier GSE57862.

\subsection*{Read alignment}

Based on the observation that many read pairs in L262 come from cDNA fragments shorter than the read length (262 bp) (Supplementary Figure 1), we identified and merged such pairs by searching for a 13 bp adapter sequence, which immediately follows the fragment sequence when the fragment is shorter than the read length. During merge, if the two reads disagreed on a base, we chose the base with a higher quality score. This resulted in a bipartition of L262 into a single-end library of 4,470,824 reads (L262S) and a paired-end library of 5,051,896 read pairs (L262P). 1,466 read pairs were discarded for being shorter than 20 bp after the  merge or having more than 10 mismatches. L262P and L262S were aligned to the reference separately and then aggregated for comparison with L75. Most of the read pairs in L75 came from fragments longer than the read length (Supplementary Figure 1), and thus we did not merge reads in L75. We aligned each library using the following tools with default settings: TopHat2 (version 2.0.5) ~\cite{Kim2013}, GSNAP (version 2012-07-20) ~\cite{Wu2010}, Bowtie2 (version 2.0.0-beta7) \cite{Langmead2012}, and STAR (version 2.3.0) \cite{Dobin2012}. For tools that required a list of known splice sites, we used the reference gene annotations from GENCODE 15 \cite{Harrow2012}.

\subsection*{Concordance analysis}

The VCF file containing positions genotyped for GM12878 was obtained from the 1000 Genomes Project (http://www.1000genomes.org). After discarding insertions, deletions (indels), and sites with incomplete genotypes, roughly 5.4M sites remained for analysis. For each dataset of aligned reads, genotyped sites covered by at least 10 reads (20-40K sites) were analyzed for concordance between the DNA-seq based genotype from the VCF file and the observed genotype from the RNA-seq reads. For each genotyped site $i$, let $n_i$ be the total number of reads covering the site, and $x_i$ be the number of {\it mismatched} reads, which is defined as reads that contain an allele not supported by the known genotype. Under the null hypothesis that there is no mapping error, the number of mismatched reads follows $\text{Binomial}(n_i,\epsilon)$, where $\epsilon$ is the per-base sequencing error rate. The {\it discordance $p$-value} is defined as $P(X \ge x_i)$, where $X\sim \text{Binomial}(n_i,\epsilon)$. Intuitively, the more significant this $p$-value is, the more confident we are that mapping error affected RNA-seq reads at the corresponding site. We estimated $\epsilon$ separately for each dataset by dividing the total number of mismatches across all sites by the total number of aligned bases (Supplementary Figure 3). Then, for each mapper, we compared the distribution of discordance $p$-values between L262 and L75 using a one-sided Wilcoxon rank-sum test. We controlled for differences in statistical power and technical variance by stratifying the sites into four groups based on read depth and performing a separate comparison for each bin: 1) 10-99, 2) 100-499, 3) 500-999, and 4) $\ge$1000 reads. The read depth distribution was highly similar across read lengths for each bin (Supplementary Figure 4).

\subsection*{Mapping reads to genomic features}

To assign each read to a genomic feature --- gene or mRNA isoform --- for the purposes of quantifying gene abundance or analyzing isoform disambiguation, we constructed a feature set based on the annotations from GENCODE 15 \cite{Harrow2012} for each locus in the reference genome representing the features that contain the corresponding nucleotide. Then, for each aligned read, we took the intersection of all nonempty feature sets associated with the bases covered by the read. Only considering the nonempty sets allows the reads to have bases that are not explained by the reference annotations, as long as the reads contain other informative bases. If the resulting set of features (which we refer to as \emph{consistent} features) has a single element, it implies that the read pair is unambiguously mapped to a feature. For mRNA isoforms, we also incorporated information about splice junctions to obtain a more precise feature mapping; when a read alignment included one or more splicing events, we only considered the set of isoforms whose splicing patterns agree with the read.

\subsection*{Detection of allele-specific expression patterns}

Phased genotypes for NA12878 were inferred by family relationships using Ped-IBD \cite{Li2010a}, while discarding sites with Mendelian inconsistencies. For each aligned read, we examined all heterozygous sites that appear in the read and assigned the read to either the paternal or maternal haplotype only if all heterozygous sites have the corresponding alleles. Gene-level ASE was assessed with a binomial test based on the number of maternal and paternal reads assigned to a given gene, assuming a balanced distribution as the null. We detected ASAS by testing for differential usage of individual exon blocks using a procedure developed in \cite{Salzman2011}. First, the exonic regions of each gene were divided into ``exon blocks" such that each block does not contain any of the exon boundaries of known mRNA isoforms. Then we tested whether each exon block exhibits a differential usage pattern using a chi-square test on a 2-by-2 contingency table where the rows indicate whether a read is maternal or paternal and the columns indicate whether a read overlaps with the exon block of interest (as illustrated in Supplementary Figure 5). We excluded human leukocyte antigen (HLA) complex genes from the analysis to avoid cases with possibly incorrect genotypes, and also disregarded genes with low total read counts (less than 20 for ASE and 40 for ASAS). Significant cases were called at 10\% false discovery rate using the Benjamini and Hochberg procedure \cite{Benjamini1995}. 

\subsection*{Calculation of sequencing costs}

We estimated the sequencing costs for L262 and L75 as follows. For L262, a 500 cycle MiSeq sequencing cartridge cost \$880, which yielded a total of 9,524,186 read pairs ($\sim$4.99 billion bases). For L75, Centrillion Biosciences charged us \$2400 for a single 2x75 lane, which we used to sequence six samples. Thus, a single L75 sample alone costs roughly \$400, which yielded 25,933,924 read pairs ($\sim$3.89 billion bases). The resulting cost estimate per million bases is \$$0.176$ for L262 and \$$0.103$ for L75. The cost of library preparation (approximately \$60 for each library) is not included in the estimate.

\section*{Results}

\subsection*{Performance of read-mapping tools}
\begin{figure}[!ht]
\begin{center}
\includegraphics[width=5in]{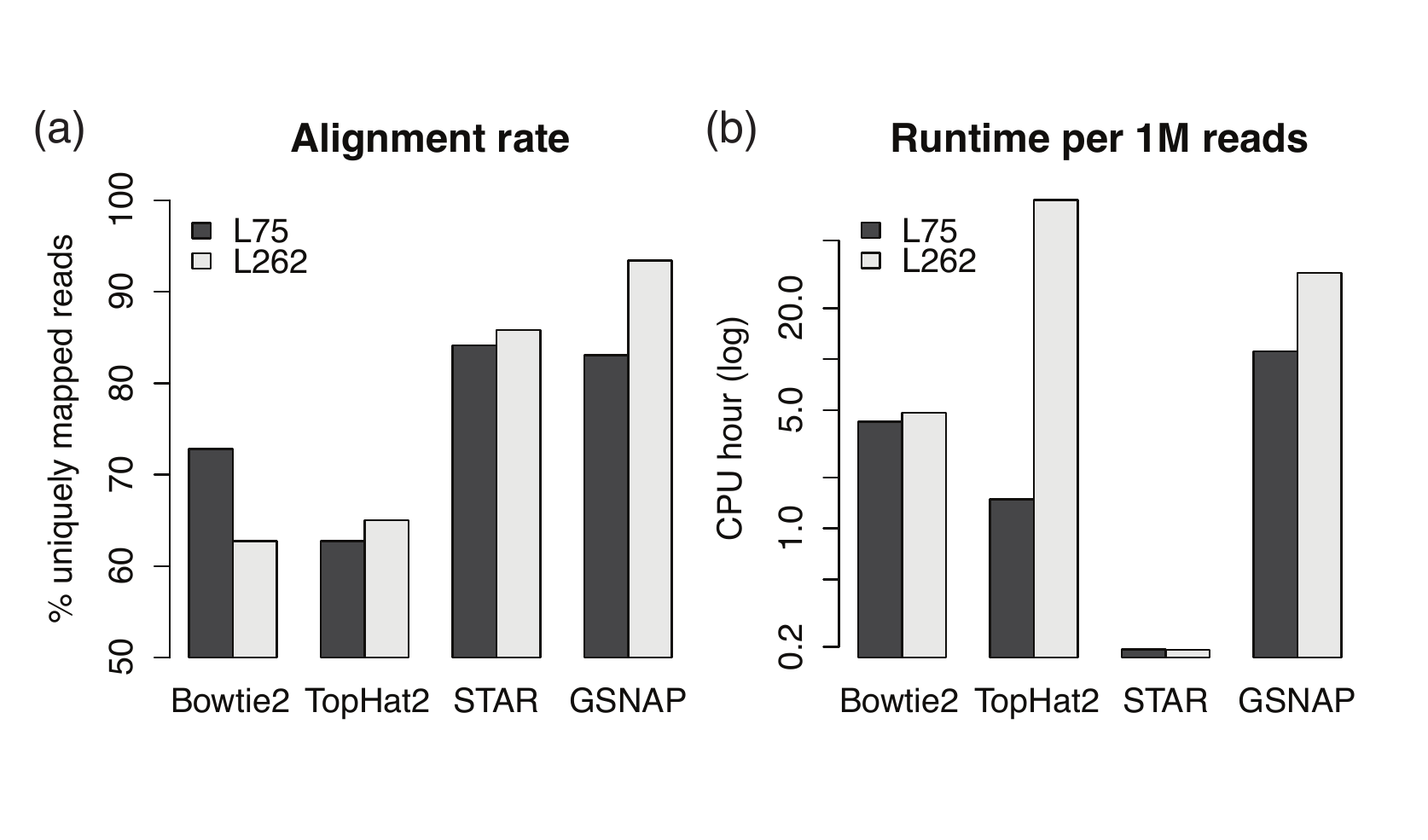}
\end{center}
\caption{
{\bf The effect of read length on read-mapping performance.}
We compared the percentage of reads uniquely mapped (a) and the average runtime per a million reads (b) of Bowtie2, TopHat2, STAR, and GSNAP on L262 and L75. Only the splice-mappers -- TopHat2, STAR, and GSNAP -- achieved higher alignment rates for L262 compared to L75. The increase in runtime going from L75 to L262 varied greatly across the mappers, TopHat2 being the most sensitive among the four mappers tested.
}
\label{fig:mapping}
\end{figure}

To analyze the impact of read length on the performance of currently available read-mapping tools, both L262 and L75 were aligned to the reference genome (GRCh37/hg19) using Bowtie2, TopHat2, STAR, and GSNAP (Methods). We emphasize that our goal is not to compare the performance of different methods on a particular dataset; rather, we aim to analyze how different methods scale to longer reads. Figure \ref{fig:mapping}a shows the percentage of reads uniquely mapped to the reference genome. For tools that adequately handle reads crossing over one or more splice junctions (TopHat2, STAR, and GSNAP), the alignment rate of L262 was consistently higher than that of L75, suggesting that longer reads can improve overall alignment rate. The difference was most dramatic for GSNAP, for which the rate was higher by 10.39 percentage points (pp) for L262 than for L75. In contrast, Bowtie2 suffered greatly for L262, achieving a rate 10.07 pp lower than that of L75. This implies that the appropriate handling of splice junctions is a crucial requirement for a mapping tool to be effectively applied to libraries with longer read lengths, likely because longer reads have a higher chance of crossing over one or more splice junctions (Supplementary Figure 6). To illustrate, in the GSNAP output, 50.18\% of uniquely mapped reads in L262 contain at least one splice junction, while the same is true for only 38.24\% of uniquely mapped reads in L75. Higher alignment rate of longer reads, however, comes at a computational cost: the execution time of splice junction-aware mapping tools was much longer for L262 than for L75 (Figure \ref{fig:mapping}b). Most notably, TopHat2's average runtime per read was approximately 58 times longer for L262, suggesting that TopHat2's underlying alignment algorithm is more sensitive to read length than the other tools we tested. STAR was the fastest tool for both libraries, but it used a considerably larger amount of memory than others (data not shown). These results demonstrate varying degrees of scalability of currently available read-mapping tools. 

\subsection*{RNA and DNA genotype concordance}
\begin{figure}[!ht]
\begin{center}
\includegraphics[width=6in]{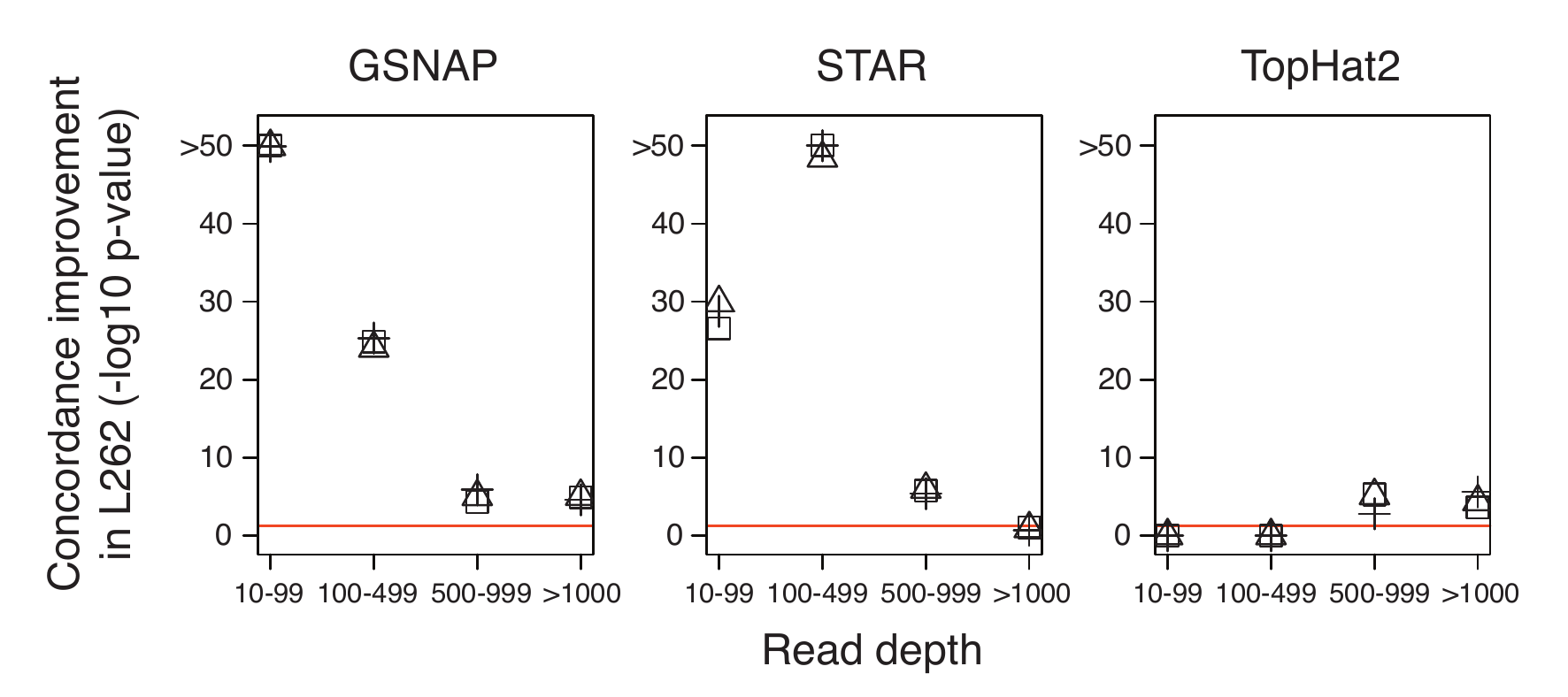}
\end{center}
\caption{
{\bf The effect of read length on RNA genotype concordance.} Results of comparisons of discordance p-values (-log10 transformed) between L262 and base-subsampled L75 stratified by mapper and read depth. Sites were stratified into four groups based on read depth in each dataset: 1) 10-99, 2) 100-499, 3) 500-999, and 4) $\ge$1000 reads. For each mapper, at each of the four read depths, a one-sided Wilcoxon rank-sum test was conducted to determine whether the $p$-values for the L262 dataset were significantly higher compared to L75. The red horizontal line represents a significance level of $\alpha =0.05$. Each symbol represents an independently subsampled data for L75.
}
\label{fig:snp_conc}
\end{figure}

The higher proportion of mapped reads observed in L262 could in theory reflect an undesirable increase in uncertain, \emph{incorrectly} mapped reads.  To address this possibility, we evaluate evidence of read mismapping as indicated by genotype calls from RNA-seq reads. We sought to compare the effect of read length on RNA/DNA genotype concordance by analyzing separately the concordance of SNPs and indels. For SNPs, using L75 and L262 mappings with STAR, GSNAP, and TopHat2, we calculated a \emph{discordance $p$-value} at each genotyped site, which represents the level of deviation from the expected genotype due to inaccurate mapping, while controlling for dataset-specific sequencing error rate (Methods). The distribution of discordance $p$-values was significantly higher (more concordant) in L262 across all read depth bins for GSNAP and in all but one bin for STAR (Figure \ref{fig:snp_conc}). This suggests that, in addition to an increase in the \emph{proportion} of uniquely aligned reads, the longer read protocol resulted in a lower rate of mapping errors. For TopHat2, the improvement in concordance for long reads was not as clear, but we suspect this is in part due to the lack of statistical power caused by low alignment rate. Next, we sought to assess whether read length influences genotype concordance estimates for indels of varying size (Supplementary Figure 8). In general, we did not observe major effects of read length, and STAR and GSNAP perform reasonably well in mapping indels as large as 20 bp in length. STAR appears to perform equally well in mapping insertions and deletions, performing best with the long reads whereas GSNAP appears to have little gain and an overall bias in mapping deletions. TopHat2, on the other hand, is very limited in its ability to call indels, maxing out in its ability to call an indel at 2 bp in length. We hypothesize that the lack of improvement of indel genotype concordance for longer reads across all methods indicates that current limitations are predominantly algorithmic. Based on the mapping rate, the GSNAP output was used for the remainder of the study.

\subsection*{Quantification of gene and isoform abundance}
\begin{figure}[!ht]
\begin{center}
\includegraphics[width=6.0in]{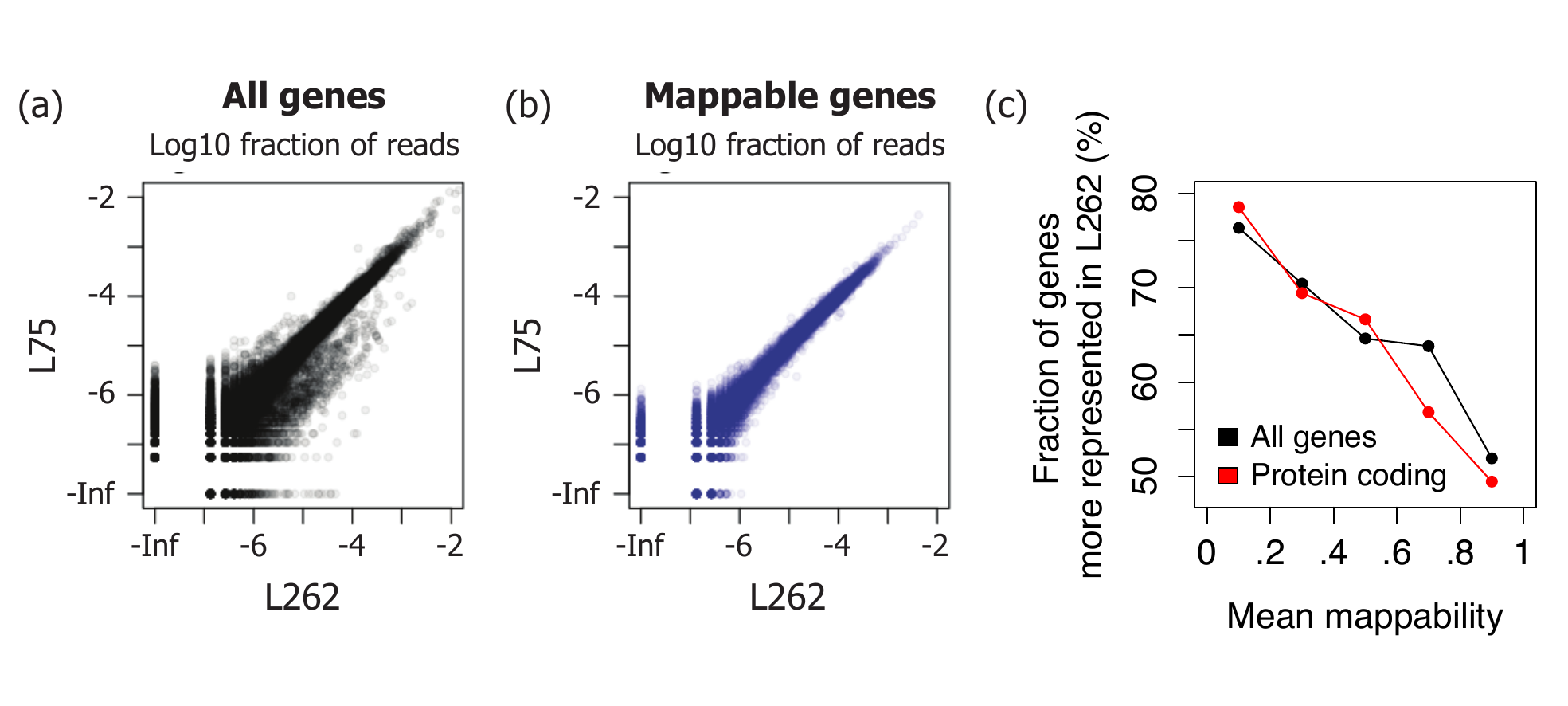}
\end{center}
\caption{
{\bf Most of the difference in gene quantification arise from poorly mappable genes.}
(a, b) Log scatter plots of fraction of reads mapped to each gene between L262 and L75. Only plotting genes that are not pseudogenes and have perfect mappability scores leads to a scatter plot with near perfect correlation. (c) Genes that are observed in both libraries and are at least 500 bp long were divided into five groups according to the mean mappability score, which is obtained by summarizing the 75 bp mappability score track on UCSC Genome Browser. For each group, we computed the fraction of genes that had a higher proportion of reads in L262 than in L75. The groups corresponding to lower mappability scores showed higher proportion of genes more represented in L262. The same trend was observed even when we restricted our analysis to protein coding genes.
}
\label{fig:gene_quant}
\end{figure}

To analyze the effect of read length on the quantification of gene abundance, we calculated the number of reads that unambiguously mapped to each gene (Methods) and compared the relative abundance between L262 and L75. Overall, the normalized read counts were highly correlated between the two libraries (r = 0.94), but we found an excess of genes that are more represented in L262 (Figure \ref{fig:gene_quant}a). For instance, among 13,969 genes observed in both libraries (with a fraction of reads $> 10^{-6}$), 139 genes displayed more than tenfold increase in L262, whereas only one gene (a pseudogene named RP5-857K21.11) showed more than tenfold decrease in L262.  Among the 139 genes that were more represented in L262, 130 were pseudogenes, which suggested that longer reads are more effective at capturing regions of high sequence similarity with other regions in the genome. In fact, 3.64\% of uniquely mapped reads in L262 are from pseudogenes, which is considerably larger than 1.25\% for L75 (Supplementary Figure 7). Among the few protein coding genes that displayed more than tenfold increase in L262 were GLUD2 and NACA2, which are known to be highly homologous to GLUD1 and NACA, respectively. To further test whether long reads are more effective at capturing non-unique regions, we used the mappability score obtained from the UCSC Genome Browser \cite{Meyer2013} as a measure of sequence uniqueness. Sorting the genes according to average mappability score revealed that genes with low mappability score tend to be more represented in L262 than in L75 (Figure \ref{fig:gene_quant}c). The same trend could be observed when we restricted our analysis to protein coding genes. Leaving out pseudogenes and genes with an average mappability score of less than one left 25,597 genes (out of 56,680), and the correlation between normalized read counts of the two libraries for this subset was considerably higher (r = 0.97, Figure \ref{fig:gene_quant}b).

\begin{figure}[!ht]
\begin{center}
\includegraphics[width=4.5in]{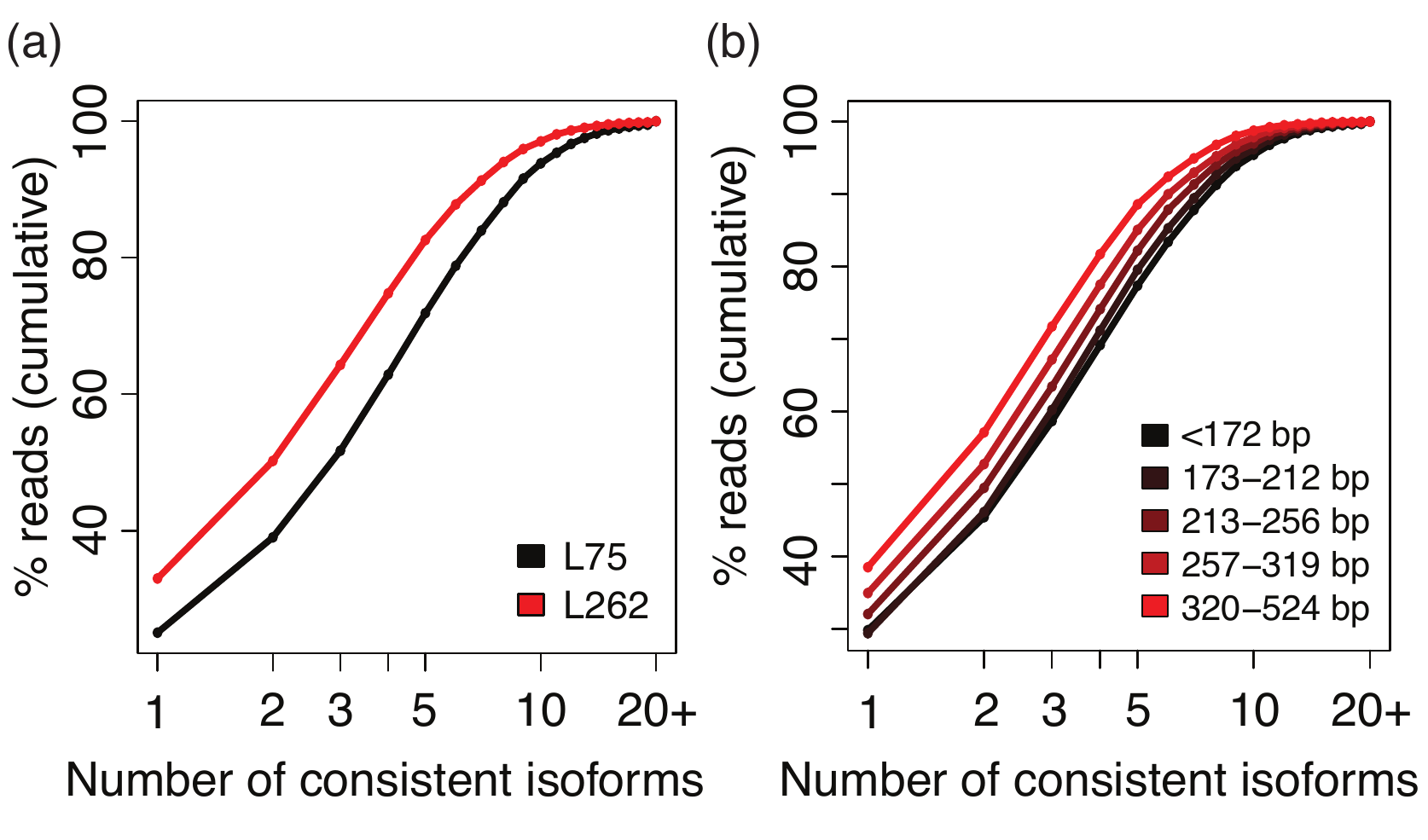}
\end{center}
\caption{
{\bf Longer reads are consistent with fewer number of mRNA isoforms.}
If all the bases of a read that maps to a gene are part of the same isoform, this common isoform (which does not have to be unique) is said to be consistent with the read. We compared the cumulative distribution of the number of consistent isoforms of reads in L75 versus reads in L262 (a) and across read groups within L262 (b) divided according to the number of reference bases covered by the alignment.
}
\label{fig:isoform_disamb}
\end{figure}

We next asked how much more effective long reads are at disambiguating mRNA isoforms. For each read in the library, we calculated the number of isoforms that can be assigned to any read (Methods) and compared its distribution between the two libraries. Overall, the reads in L262 were assigned to fewer different isoforms than those in L75 (Figure \ref{fig:isoform_disamb}a). In particular, 32.98\% of reads in L262 that were assigned to at least one isoform were unambiguously assigned to a single isoform, which is 7.94\% higher than 25.04\% observed for L75. This agrees with intuition that longer reads have increased likelihood of crossing a splice junction or containing informative bases that allow disambiguation among candidate isoforms.  Furthermore, the same pattern could be observed within a library; by stratifying the reads in L262 according to their \emph{effective alignment length} (the number of bases in the reference covered by an aligned read pair; distribution shown in Supplementary Figure 2), we observed that reads with longer alignments tend to have fewer matching isoforms indicating better specificity of the originating transcript (Figure \ref{fig:isoform_disamb}b). In fact, if we only take the subset of 104,910 read pairs in L262 with effective alignment length of 524 bp (the longest possible), the unambiguous mapping rate increases to 40.96\%.

\begin{figure}[!ht]
\begin{center}
\includegraphics[width=5.8in]{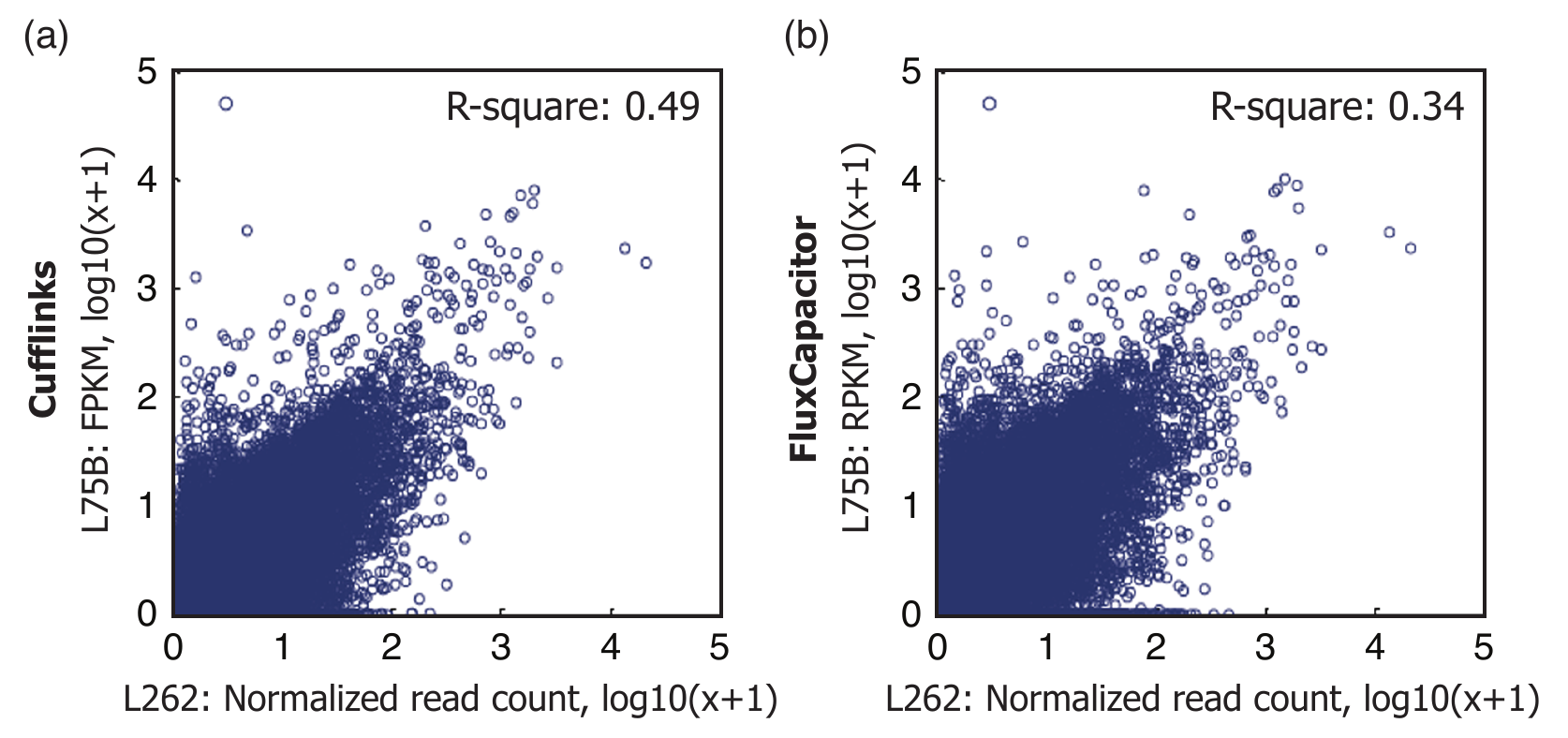}
\end{center}
\caption{
{\bf Comparison of Cufflinks and FluxCapacitor on transcript quantification.} 
We used transcript counts from L262 to assess performance of (a) Cufflinks and (b) FluxCapacitor on L75. Cufflinks and FluxCapacitor were respectively applied on base-subsampled L75 to generate FPKM/RPKM counts. On L262, we only counted reads that unambiguously mapped to each transcript, and divided each by the number of bases that are unique to that transcript. We restricted our analysis to 41,715 protein coding transcripts with at least 10 unique bases. Using this approach, we identified that Cufflinks has significantly better correlation to the long-read data than FluxCapacitor ($p$-value $<10^{-163}$).
}
\label{fig:iso_compare}
\end{figure}

As longer read data provides the opportunity to test the efficacy of transcript abundance methods, we analyzed the relative ability of both Cufflinks and FluxCapacitor to infer transcript abundances from L75 data. Here, we used the L75 data mapped with GSNAP and ran both methods. We then compared the correlation of estimated transcript abundances to those directly measured in the L262 data. Transcript abundances from L262 were obtained by identifying reads which unambiguously mapped to single transcripts, summarizing counts for each transcript, and further dividing each by the number of bases that are unique to that transcript. The latter provides a means to account for the fact that counts are biased by the effective proportion of the transcript length that is unique. Using this straightforward approach to evaluate this class of tools, we identified better correlation with Cufflinks ($R^{2} = 0.49$) than FluxCapacitor ($R^{2} = 0.34$; Figure \ref{fig:iso_compare}).   

\subsection*{Detection of allele-specific expression patterns}
\begin{figure}[!ht]
\begin{center}
\includegraphics[width=6.0in]{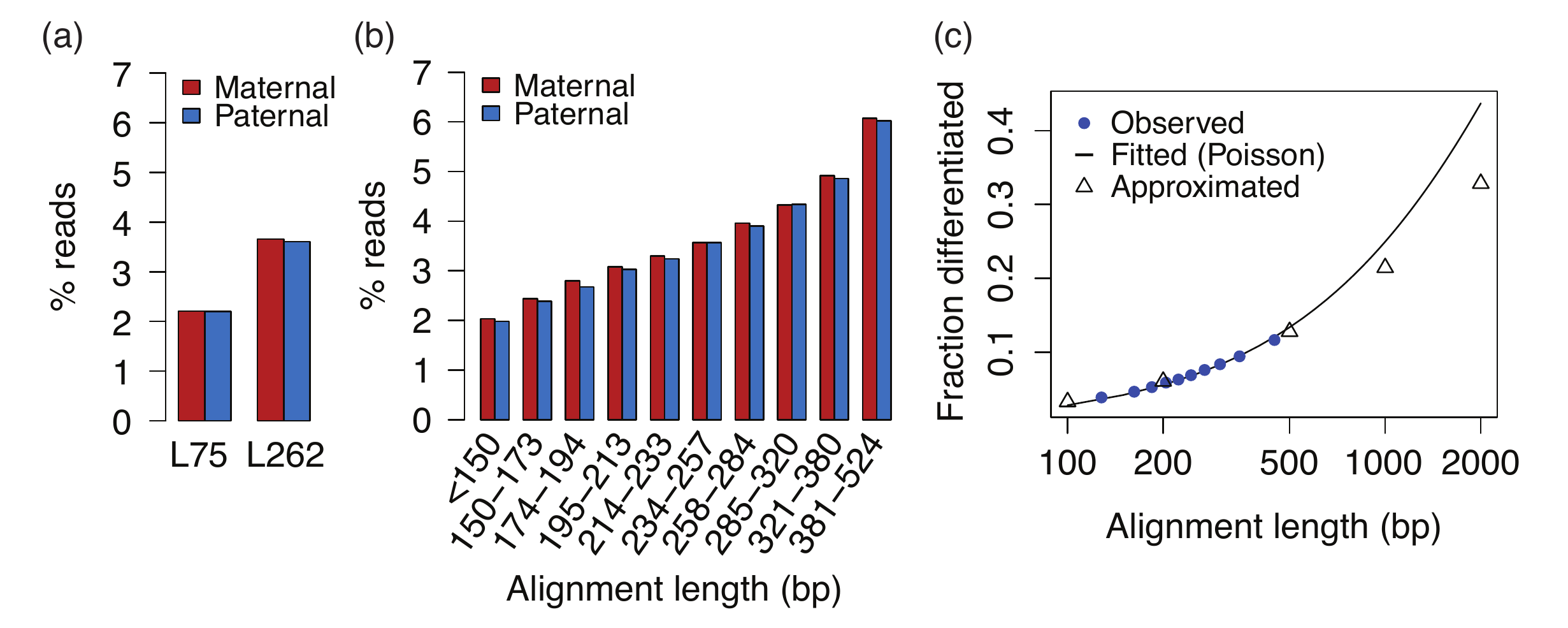}
\end{center}
\caption{
{\bf Longer reads have a higher chance of containing a heterozygous site that allows us to differentiate the haplotype of origin. }
We compared the proportion of reads unambiguously assigned to maternal or paternal haplotypes between L75 and L262 (a) and within L262 (b), where reads in L262 were grouped by the number of reference bases covered by the alignment of (possibly overlapping) paired-end reads. The observed data in (b) were extrapolated using a poisson model to even longer read lengths. To provide additional support for the extrapolation, we directly approximated the proportion of reads containing at least one heterozygous site based on the GENCODE annotation and the genotype of this individual, assuming that the true gene expression levels are the same as those observed in L262, that all mRNA isoforms of a gene are uniformly expressed, and that each starting location of a transcript is equally likely to be included in the library. We only considered reads that mapped to autosomal chromosomes. Our approximation agreed well with both the observed data and the extrapolation.
}
\label{fig:haplodiff}
\end{figure}

Studies of allele-specific expression from RNA-seq data require differentiating between alleles, which is possible for reads over one or more heterozygous loci.  Given full haplotype information for GM12878, we evaluated the fractions for which the parental haplotype in unambiguous (Methods). We were able to assign 7.01\% of the reads in L262 to either maternal or paternal haplotype as opposed to 4.26\% for L75 (Figure \ref{fig:haplodiff}a). This increased rate of haplotype differentiation for longer reads could also be seen within L262 when the reads were stratified by effective alignment length (Figure \ref{fig:haplodiff}b). Noting that the number of heterozygous sites in a given interval approximately follows a Poisson distribution, we fit a Poisson model to calculate the expected proportion of allele-specific reads for libraries with a read length greater than 262 (Figure \ref{fig:haplodiff}c). Our fitted model predicts that approximately a quarter of the reads that cover 1,000 bases in the reference will contain at least one heterozygous site.

Next, to evaluate the extent to which reference allele bias in read-mapping affects the detection of allele-specific patterns in our data, we ran the following simulation. Given the reads we classified as either paternal or maternal, we flipped the bases at heterozygous sites. We then remapped the modified reads and calculated, for each gene, the proportion of reads that mapped to the same location as before, which we refer to as {\it mapping retention rate}. When we ran this procedure for both L262 and L75, we found that L262 leads to significantly higher mapping retention rate overall (Supplementary Figure 9). The fraction of genes with retention rate lower than 0.9 was 19.27\% for L75, while it was only 7.09\% for L262. This suggests that allelic imbalance estimates measured by long reads are more reliable than those from short reads, and specifically less subject to false positives arising from mapping artifacts. For the remaining parts of this study, we disregarded genes with mapping retention rate lower than 0.9 for each library to avoid calling allelic effects as significant that may only reflect mapping errors.

\begin{figure}[!ht]
\begin{center}
\includegraphics[width=6in]{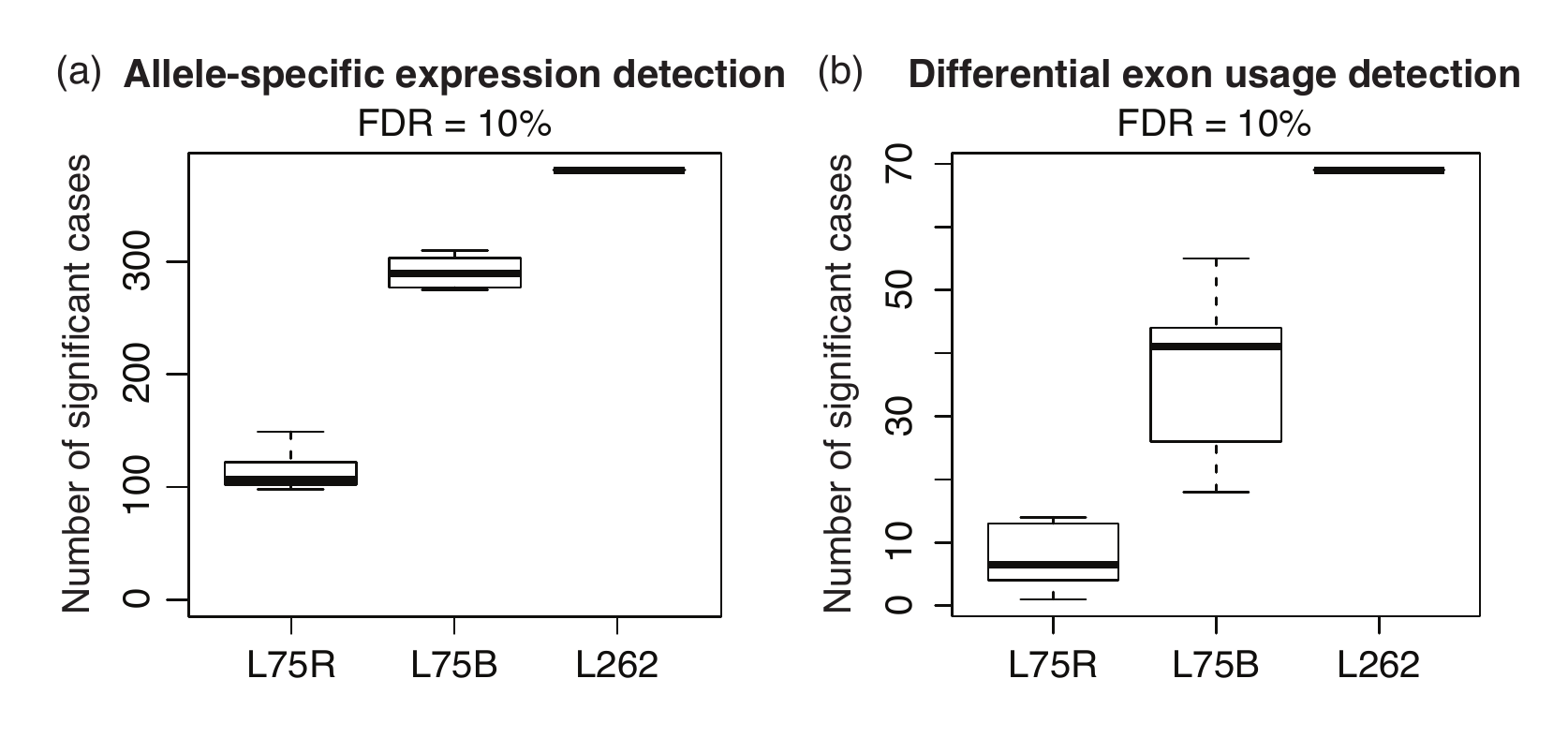}
\end{center}
\caption{
{\bf Longer reads enable more effective detection of allele-specific patterns.}
To make a fair comparison, L75 was randomly subsampled ($n=10$) down to the same number of bases (L75B) and the same number of reads (L75R) as L262. We ran the subsampled libraries and L262 through the pipeline for detecting allele-specific gene expression and allele-specific exon inclusion-exclusion patterns. The figures above show the number of significant cases we called at FDR = 10\% using the Benjamini and Hochberg procedure \cite{Benjamini1995}.
}
\label{fig:allele_specific}
\end{figure}

Because longer read length leads to more reads with heterozygous sites and fewer genes marked as untestable by the mapping retention rate filter, L262 is expected to be statistically more powerful in discovering allele-specific expression patterns. Indeed, when we checked the number of significant cases of gene-level ASE (Methods) at FDR = 10\%, we got a considerably larger number with L262 than L75, where the latter was randomly subsampled in two ways for comparison by matching either the number of bases or the number of reads (Figure \ref{fig:allele_specific}a). We also observed fewer significant cases (180 at FDR = 10\%) in a simulated short-read library constructed by truncating each read in L262 to 75 bp. Interestingly, 206 out of 381 significant cases found with L262 were not called significant by any of the subsampled L75 libraries, and among those, 24 did not pass the mapping retention rate filter in L75. One potential explanation for cases that are only detected in L262 is related to mappability: the mean mappability scores of genes with significant ASE in L262 that were not detected in L75 were significantly lower than that of genes called significant in both (Wilcoxon rank-sum $p$-value:  $3.70\times 10^{-4}$). To further validate significant ASE called with L262, we cross-referenced our significant results with a list of known eQTLs from a large population based study \cite{Montgomery2010}. 46 out of 3,328 tested genes had a heterozygous site at a known eQTL SNP for the gene, and 12 of those cases appeared in the final set of 381 significant ASEs, showing a considerable enrichment ($\chi ^2$ $p$-value = $0.018$). These results are further confirmed using a standard, per-locus ASE test (as in \cite{Montgomery2010}). In this case, we also observed a higher rate of ASE detection in L262 (Supplementary Figure 10).

\begin{figure}[!ht]
\begin{center}
\includegraphics[width=6in]{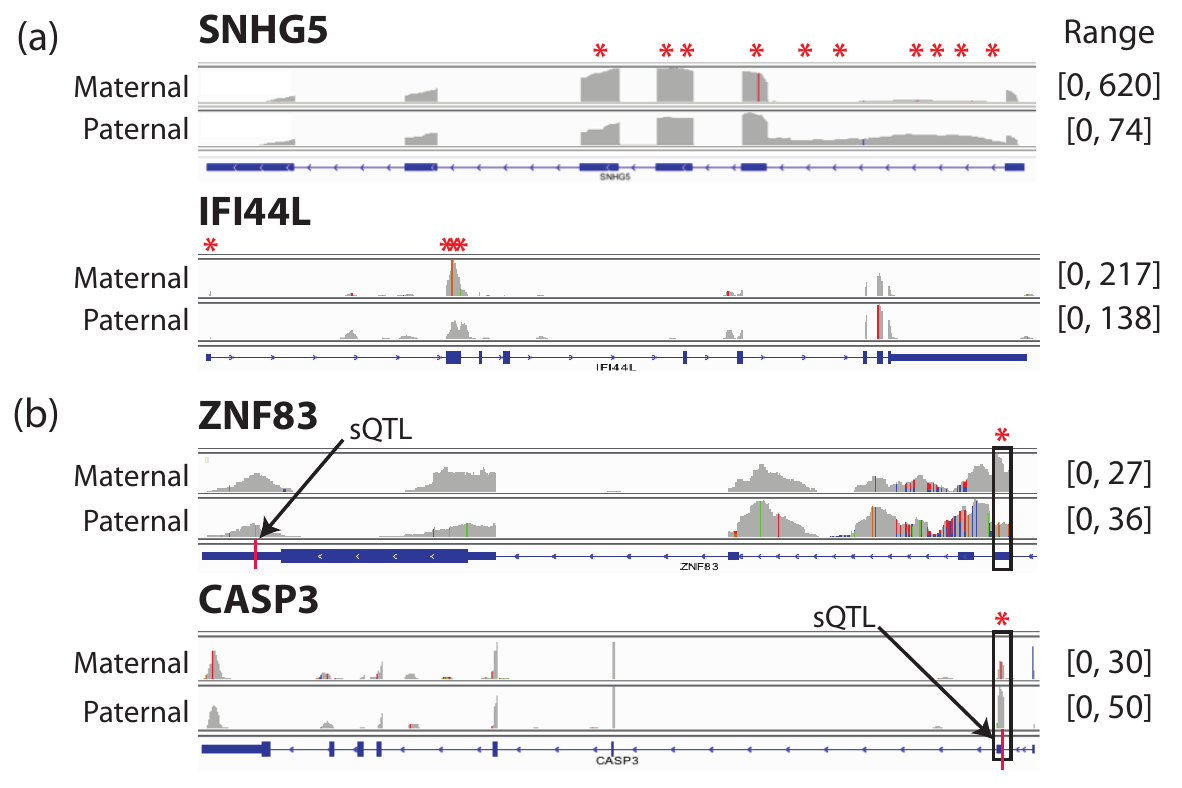}
\end{center}
\caption{
{\bf Examples of allele-specific isoform distribution detected with L262.}
Shown are the IGV \cite{Thorvaldsdottir2013} coverage plots of (a) genes corresponding to the top two significant cases (raw $p$-value $<10^{-12}$) and (b) two significant cases (FDR = 10\%) with a strong prior support by having a heterozygous site at a known sQTL from a large population-based study. The implicated sQTL and the exon that is known to be affected are noted by a red line and a black box, respectively.  Red asterisks above each plot denote blocks with significant differential inclusion-exclusion pattern (FDR = 10\%).
}
\label{fig:exonusage}
\end{figure}

We next looked for a more complex form of allelic effects, allele-specific alternative splicing, by assessing haplotype-specific differences in isoform abundance with the exon inclusion-exclusion test developed in \cite{Salzman2011} (Methods), in which the ratio of reads that support the inclusion or exclusion of an exon block is compared between the two haplotypes. At FDR = 10\%, we obtained 69 significant cases of differential exon usage with L262, which supported ASAS of 38 distinct genes. Subsampled libraries of L75 detected significantly fewer cases in general (Figure \ref{fig:allele_specific}), resulting in a median of 41 differentially used exon blocks (24 genes) for the base-subsampled L75 and 6.5 exon blocks (4.5 genes) for the read-subsampled L75. A simulated short-read library constructed by truncating each read in L262 to 75 bp also resulted in fewer discoveries (37 exon blocks, representing 25 genes). The coverage plots of the top two cases detected from L262 -- SNHG5 and IFI44L -- with raw $p$-value $<10^{-12}$ show clear haplotype-specific difference in distribution of reads across exons of the genes (Figure \ref{fig:exonusage}). Among other significant findings, ZNF83 (raw $p$-value = $1.44\times 10^{-4}$) and CASP3 (raw $p$-value = $3.22\times 10^{-4}$) are supported by prior knowledge; they contain a differentially used exon block that is known to be affected by an sQTL \cite{Battle2013} where this individual has a heterozygous genotype (Figure \ref{fig:exonusage}). CASP3 was not called significant in any of the subsampled L75 libraries.  These examples of ASAS represent likely cases of a cis-acting genetic variant altering splicing in this individual.

\subsection*{Sequencing cost comparison}

While we have shown the degree to which longer read length provides a number of key benefits for transcriptome analysis, it is important to note that it is generally more expensive. Based on actual costs, we estimated the sequencing cost per million bases of L262 to be \$$0.176$ in contrast to \$$0.103$ of L75 (Methods), which implies that one 262 bp paired-end read costs approximately the same as six 75 bp paired-end reads. The sheer number of additional short reads one can obtain in lieu of increasing read length renders some of the benefits of longer reads cost-ineffective. In particular, the number of reads unambiguously mapped to a genomic location, an mRNA isoform, or a parental haplotype would all be significantly higher if we were to use a short read dataset that is six times larger than the cost-matched long read dataset. In terms of the ability to detect allele-specific patterns, L262 was approximately \$364 more expensive than the base-subsampled L75, which resulted in 24\% fewer cases in ASE detection and 44\% fewer in ASAS detection on average. When we matched the costs between the two libraries by using all reads in L75 and 45.45\% of the reads in L262, we found that L262 actually detects 60\% fewer ASE cases and 88\% fewer ASAS cases on average than L75, which detects 418 ASE and 106 ASAS cases (Supplementary Figure 11). Nevertheless, among the cases detected by the cost-matched L262, 9\% of ASE cases and 11\% of ASAS cases were not detected by L75 because they did not pass the mapping retention rate filter. This implies that long reads, even with lower total power than short reads at the same cost, more accurately capture cases that are susceptible to mapping errors. Clearly, these results depend on the current sequencing costs; as the price of long reads becomes more comparable to that of short reads, we expect the benefits of longer read lengths to become more apparent. In addition, using longer reads fundamentally reduces the mapping bias and leads to more accurate quantification of genomic elements, which cannot be achieved by simply obtaining a larger short read dataset.

\section*{Discussion}

Longer read lengths in RNA-sequencing offer a number of advantages ultimately supporting more accurate quantification of diverse transcriptional effects, as demonstrated here through a comparative analysis of two libraries from the same cell line with different read lengths.  First, the unique alignment rate of long reads is observed to be higher, with GSNAP uniquely mapping 93.44\% of long reads in contrast to 83.05\% of short reads. Intuitively, longer sequences are less likely to occur more than once in the genome, even with the prevalence of regions sharing sequence similarity, such as gene families and pseudogenes.  In addition to the higher mapping rate, our analysis of genotype concordance suggests that longer reads are significantly less affected by mapping biases.

Next, there are differences evident in the estimates of gene expression quantified from the two libraries, which in principle could be influenced by differences in read mapping. In our study, genes with evidence of differential expression by read length were primarily genes with low mappability, and the proportion of long reads that unambiguously mapped to a pseudogene was roughly three times higher than that of short reads, with a similar observation in protein coding genes with low uniqueness. This suggests that the improved mapping of longer reads also provides more accurate estimates of gene expression. Furthermore, it reinforces that mappability is an important parameter to consider when comparing data between datasets with varying read lengths (such as for differential expression analyses).

Long reads can also reduce the ambiguity between different transcripts arising from the same gene.  Because such  transcripts share large fractions of their sequence in common, transcript-level quantification is a considerably harder task than gene-level quantification.  Only reads which span a splice junction or which include bases simply not found in a particular transcript (such as an exon excluded from one isoform) can directly disambiguate between related transcripts. Here, we demonstrated that the longer read protocol indeed allowed improved disambiguation; the proportion of exonic reads unambiguously assigned to a single isoform was 32.98\% for L262 and 25.04\% for L75. We further provide data and a straightforward means for comparing transcript abundance methods by comparing their inferences from L75 to directly measured abundances in L262; using this approach we have identified a better correlation for Cufflinks compared to FluxCapacitor.

The use of RNA-sequencing to detect allele-specific transcriptional effects offers the potential to pinpoint \textit{cis}-regulatory effects \cite{Montgomery2010, Pickrell2010}, but relies on identifying which allele (or haplotype) each read arises from.  Long reads not only increase the proportion of reads unambiguously assigned to each haplotype (7\% for L262 compared to 4\% for L75), but also diminish the susceptibility of ASE analyses to mapping biases as shown by our mapping retention rate analysis. Thus, longer reads provide a more robust quantification of allele-specific expression. We observed that the longer read dataset actually detects more significant cases than the short read dataset with the same total number of bases, and identified two particularly compelling cases of ASAS supported by a known sQTL, one of which could not be detected with short reads. Detecting such allelic effects could dramatically improve our understanding of the cis-regulation of alternative splicing within an individual.  However, initial efforts at quantifying allele-specific alternative splicing have largely relied on indirect signals such as apparent differences in insert size \cite{Montgomery2010}, or have been restricted to a small number loci with high coverage near splice junctions \cite{Lappalainen2013}.  Longer reads, which are both more likely to contain heterozygous loci and more likely to disambiguate among related transcripts and isoforms, thus offer a particularly large advantage in quantifying ASAS and identifying significant cis-regulated alternative splicing effects.  

Overall, the increasing availability of long read RNA-seq libraries offers significant advantages in analysis of the transcriptome. While the current long read sequencing technology is considerably more expensive than short read sequencing, more accurate quantification of various aspects of transcriptome due to lower mapping bias can only be achieved with longer read lengths. Where identification of alternative splicing, fusion events, and novel transcripts have traditionally required sophisticated computational methods and statistical inference, longer reads will bring us increasingly close to direct measurement of these events. Further distinguishing the benefits, long reads allow more accurate SNP genotype calls from RNA-seq data and, critically, increasingly detailed analysis of allele-specific expression. As public data increasingly includes RNA-seq data of different read lengths, understanding these advantages and caveats will lead to better tools and comparisons of heterogeneous data. 


\section*{Acknowledgments}
This work was supported by the Edward Mallinckrodt Jr. Foundation and the National Institutes of Health (R01 MH10814).


\pagebreak
\section*{Figure Legends}

\clearpage
\setcounter{figure}{0}
\renewcommand{\figurename}{Supplementary Figure}

\title{High-resolution transcriptome analysis with long-read RNA sequencing\\\vspace{1em}
\Large \textbf{Supplementary Material}}

\author{Cho, H. et al.}

\maketitle

\begin{figure}[!ht]
\begin{center}
\includegraphics[scale=.7]{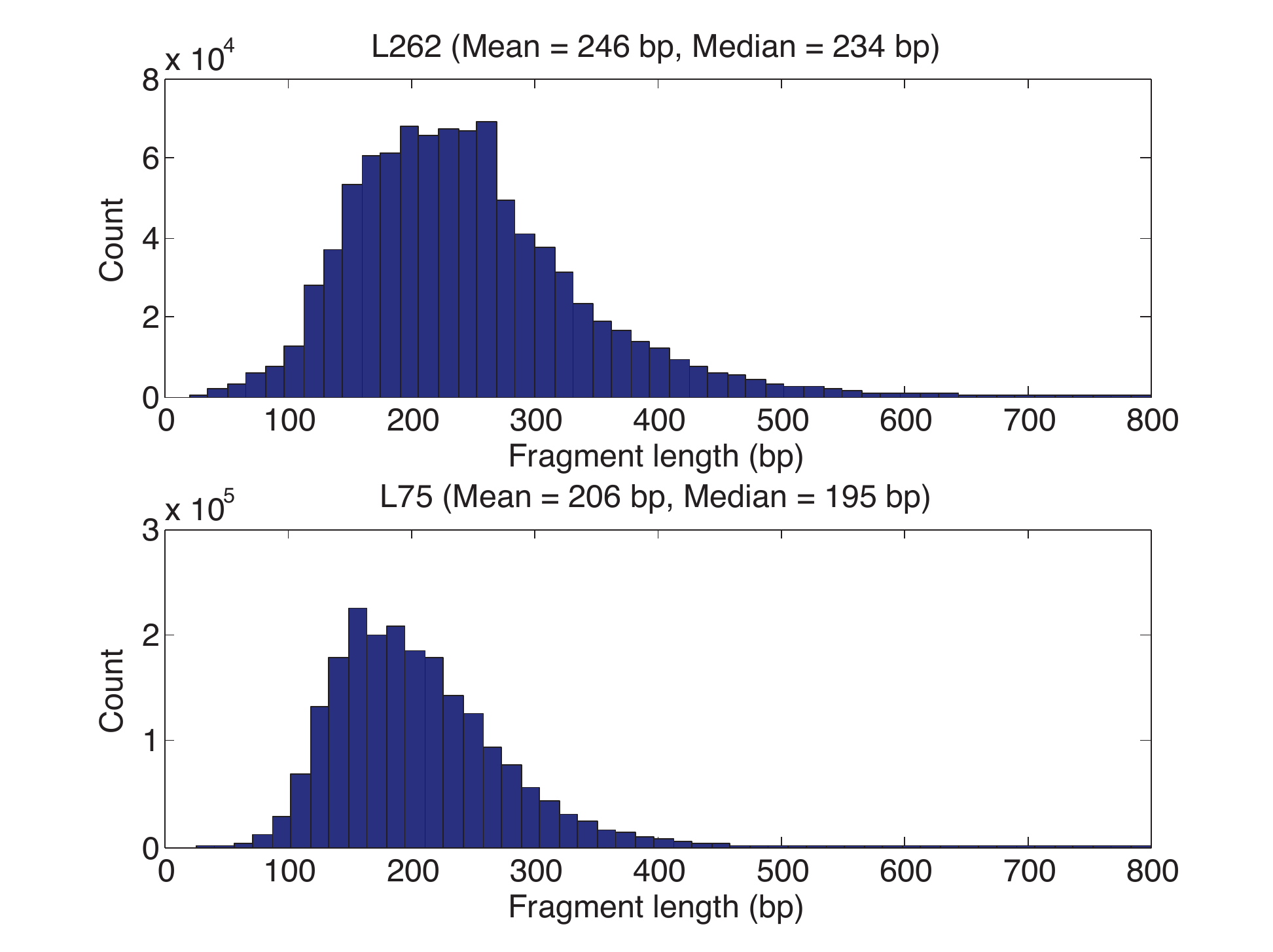}
\end{center}
\caption{
{\bf Estimated fragment length distribution.}
We used the reads that mapped to genes with exactly one transcript to estimate the fragment length distributions for L262 and L75. The read alignments produced by GSNAP were used.
}
\label{fig:fragment_len}
\end{figure}

\begin{figure}[!ht]
\begin{center}
\includegraphics[scale=.7]{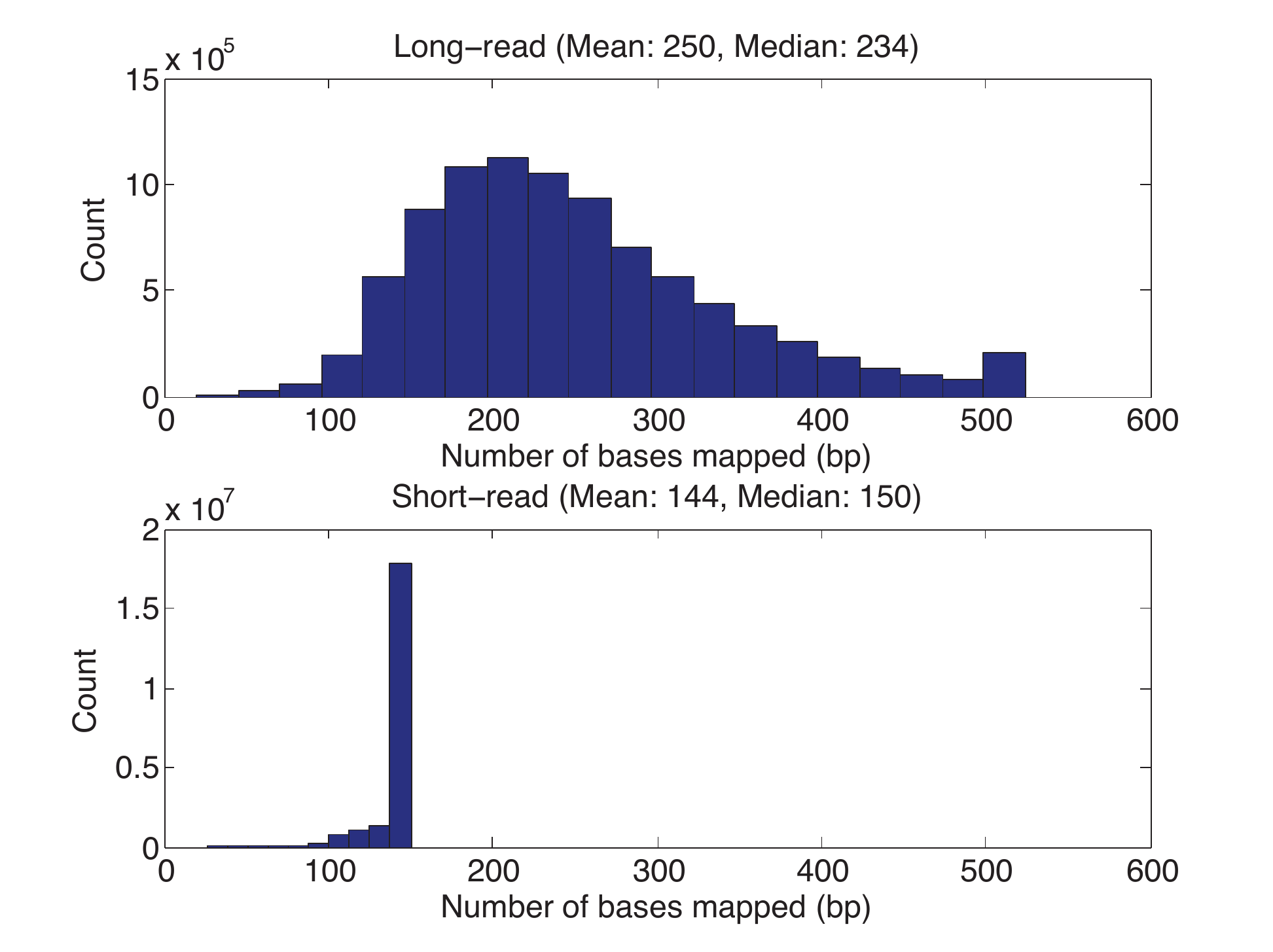}
\end{center}
\caption{
{\bf Histogram of number of reference bases covered by each read.}
As a measure of information content, we counted the number of bases in the reference genome covered by each read pair. For overlapping reads, for instance, this will be less than double the read length. From the histograms, we can see that the vast majority of reads in L75 are non-overlapping while reads in L262 show high degrees of overlap. Overall, however, L262 reads still have more information content than L75 reads.
}
\label{fig:ref_bases_covered}
\end{figure}

\begin{figure}[!ht]
\begin{center}
\includegraphics[width=5in]{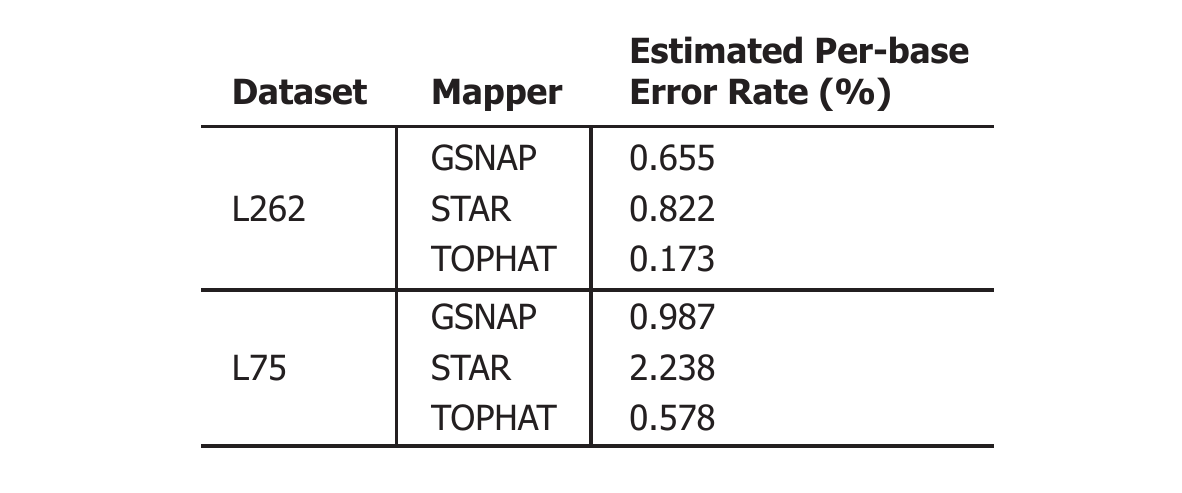}
\end{center}
\caption{
{\bf Estimated per-base sequencing error rate for each aligned dataset.}
Sequencing error rate is estimated as the total number of mismatches in the alignments divided by the total number of aligned bases.
}
\label{fig:error_rate_table}
\end{figure}

\begin{figure}[!ht]
\begin{center}
\includegraphics[width=6in]{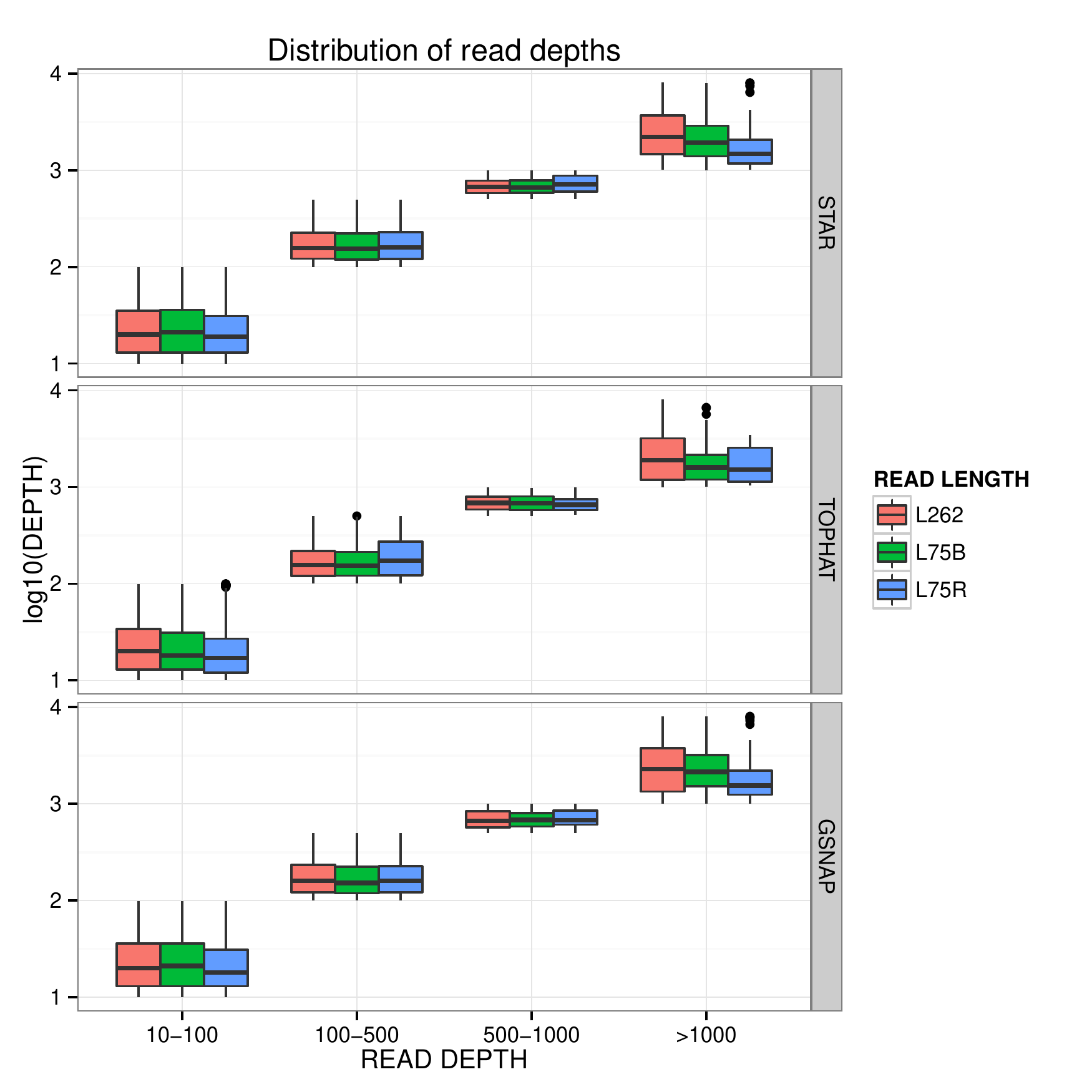}
\end{center}
\caption{{\bf Comparison of read depth distributions.} Genotyped sites were stratified into the same four read depth bins as used in the genotype concordance analysis. This plot compares the distribution of read depths within each category across mappers and read length. The distributions within each read depth category are highly similar across read lengths.}
\label{fig:depth_distribution}
\end{figure}

\begin{figure}[!ht]
\begin{center}
\includegraphics[width=6in]{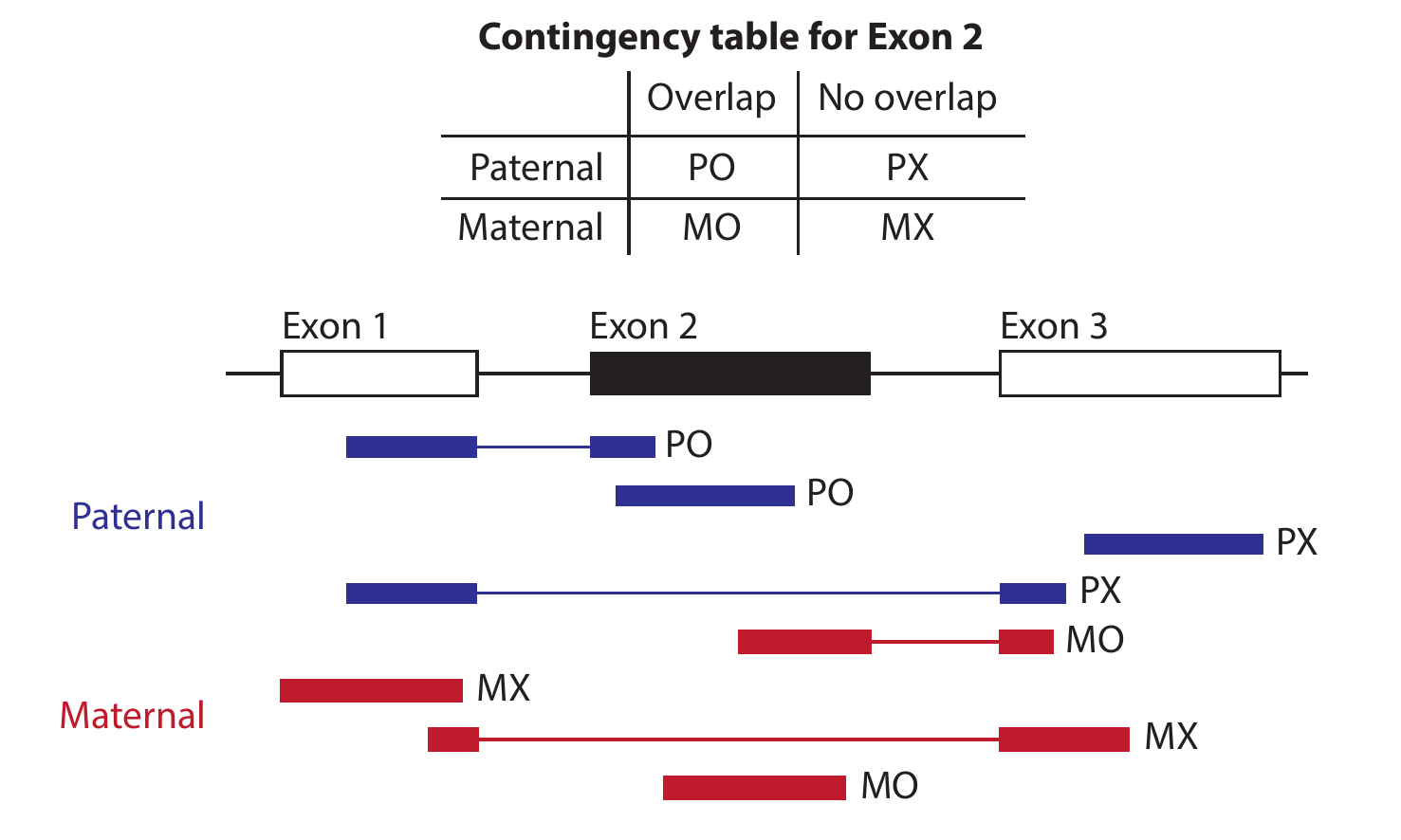}
\end{center}
\caption{{\bf Illustration of differential exon usage test.} We tested whether each exon block exhibits a differential usage pattern using a chi-square test on a 2-by-2 contingency table where the rows indicate whether a read is maternal or paternal and the columns indicate whether a read overlaps with the exon block of interest. Each example read (shown as single-end for simplicity) is marked with the cell in the contingency table it is counted towards. }
\label{fig:asas_illustration}
\end{figure}

\begin{figure}[!ht]
\begin{center}
\includegraphics[width=5in]{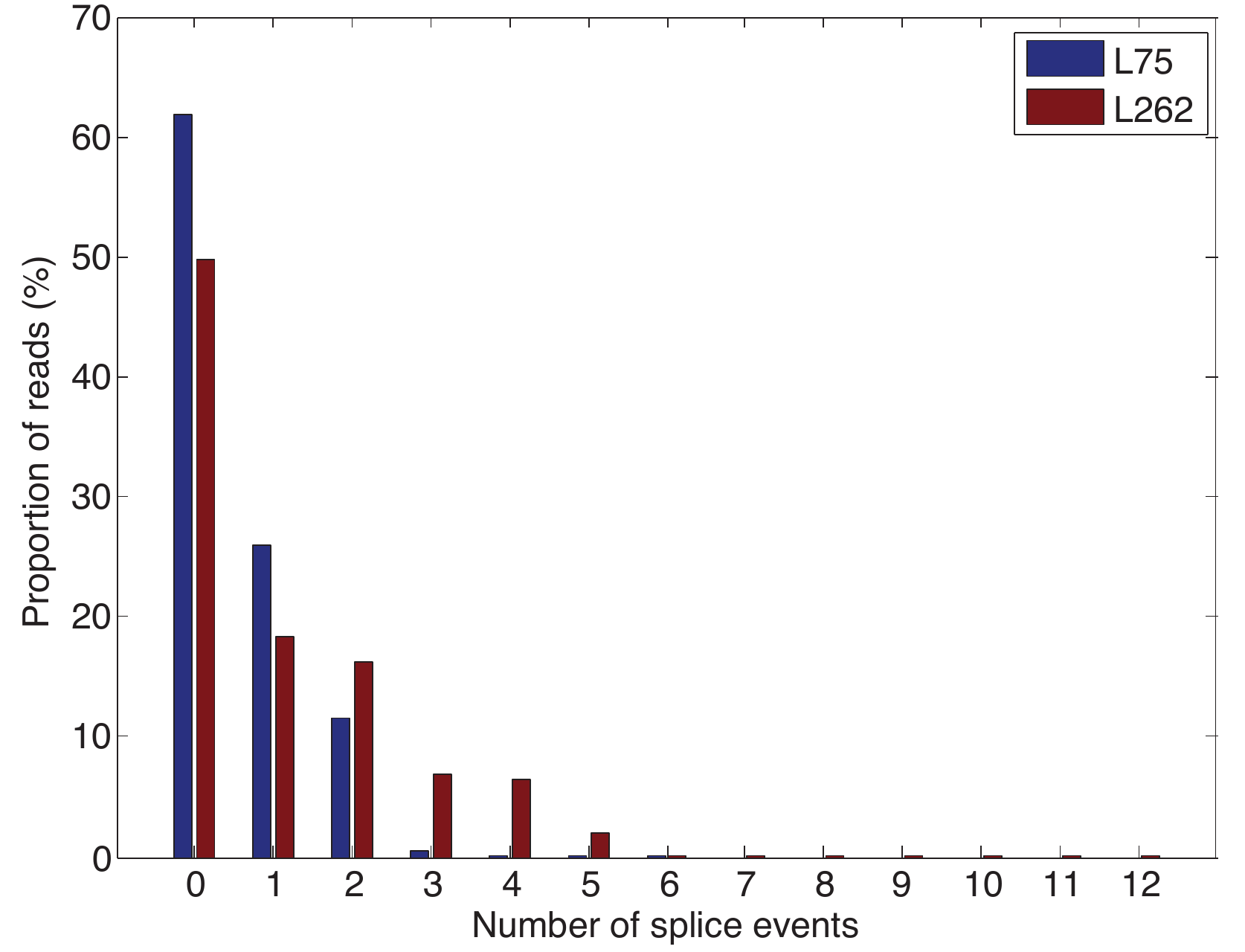}
\end{center}
\caption{
{\bf Distribution of number of splice junctions spanned by each read. }
For each read pair, we counted the number of splice events represented in the alignment from GSNAP. The bar graph shows the distribution of this number in both L262 and L75. We can see that L262 reads tend to span more splice junctions than L75.
}
\label{fig:splice_junc_count}
\end{figure}

\begin{figure}[!ht]
\begin{center}
\includegraphics[width=3in]{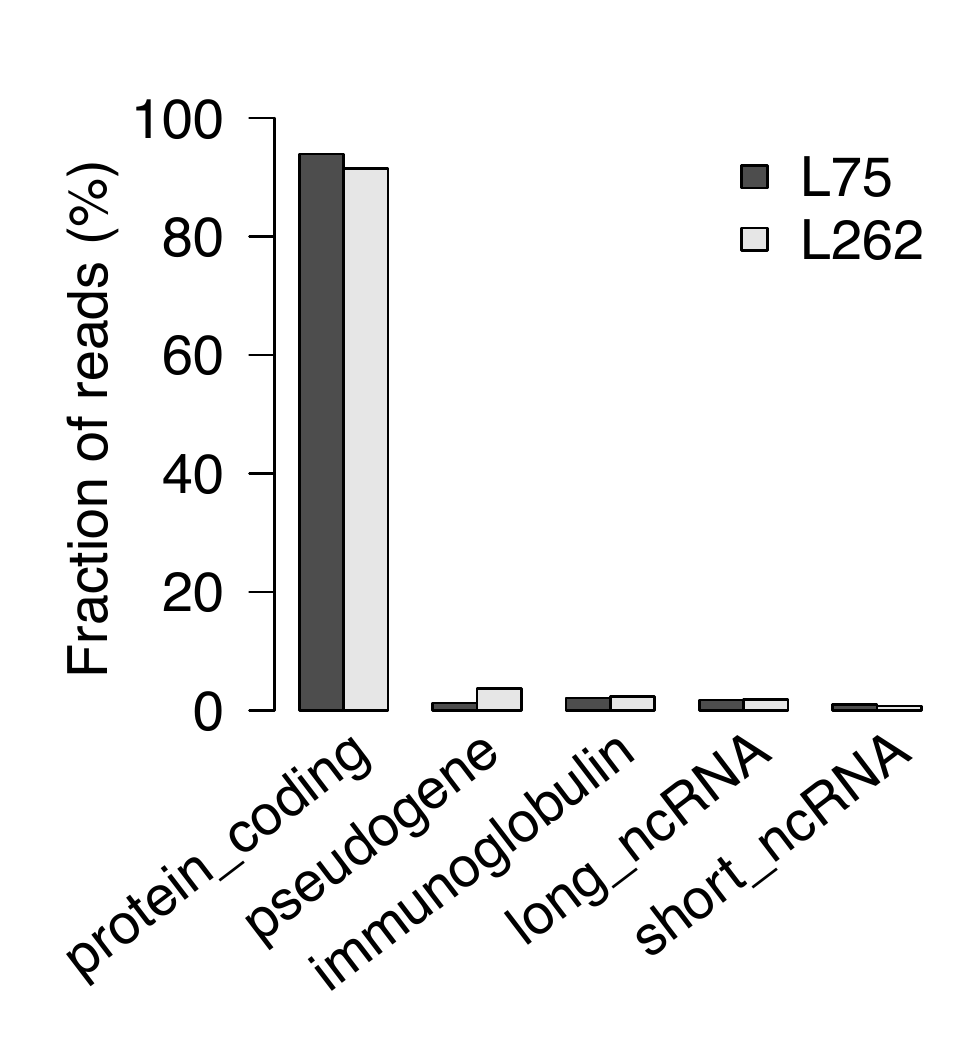}
\end{center}
\caption{
{\bf Breakdown of reads by gene category.}
The fraction of reads mapped to each of the five major gene categories in GENCODE annotation (version 15) is shown. L262 displayed a relatively high proportion of reads originating from pseudogenes compared to L75.
}
\label{fig:gene_cat}
\end{figure}

\begin{figure}[!ht]
\begin{center}
\includegraphics[width=6in]{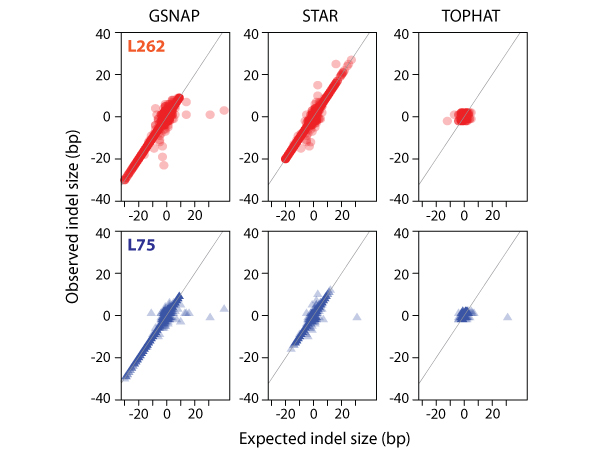}
\end{center}
\caption{
{\bf Expected indel size from DNA-seq compared to most frequently observed size from RNA-seq for L262 and L75.}
Points along the diagonal (gray line) in each plot indicate a positive correlation between most frequently observed indel size and expected size based on genotyping. In general, STAR and GSNAP perform reasonably well in mapping indels as large as 20 bp in length. Tophat, on the other hand, is very limited in its ability to call indels, maxing out in its ability to call an indel at 2 bp in length. STAR appears to perform equally well in mapping insertions and deletions, performing best with the long reads. GSNAP appears to have a bias in mapping deletions. The L75 dataset was subsampled to match the number of bases in L262 for this analysis.
}
\label{fig:indel_sizes}
\end{figure}

\begin{figure}[!ht]
\begin{center}
\includegraphics[width=4in]{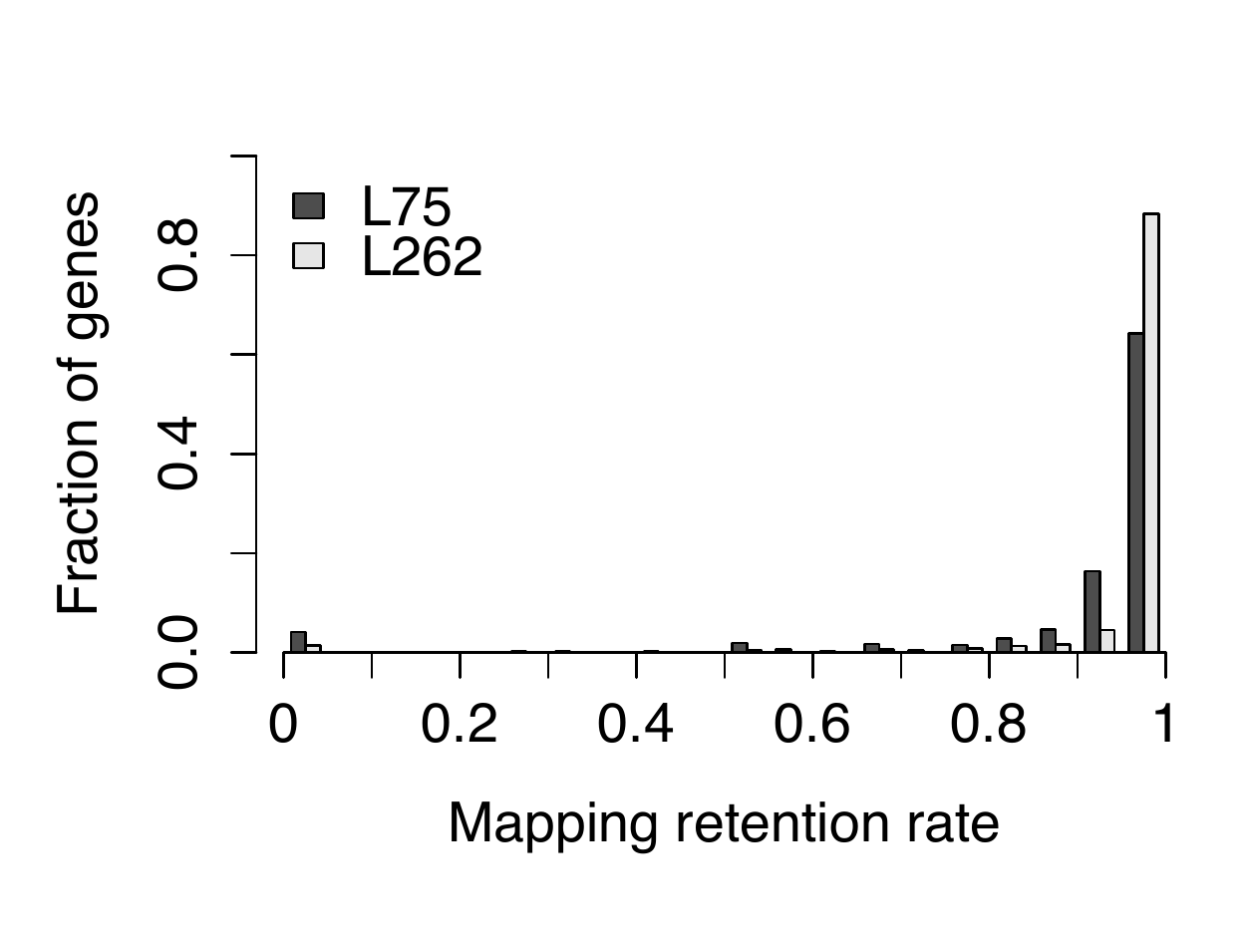}
\end{center}
\caption{
{\bf Genes quantified by longer reads display higher mapping retention rate.}
Mapping retention rate is a simulation-based metric that is inversely correlated with the severity of allelic bias in allele-specific quantification. It is computed for each gene as the proportion of reads that map to the same location after flipping all bases that align with heterozygous sites in the reference so that the new bases are the alleles from the other parental haplotype. The above histogram compares the distribution of mapping retention rates in L75 versus L262.
}
\label{fig:ret_rate}
\end{figure}

\begin{figure}[!ht]
\begin{center}
\includegraphics[width=6in]{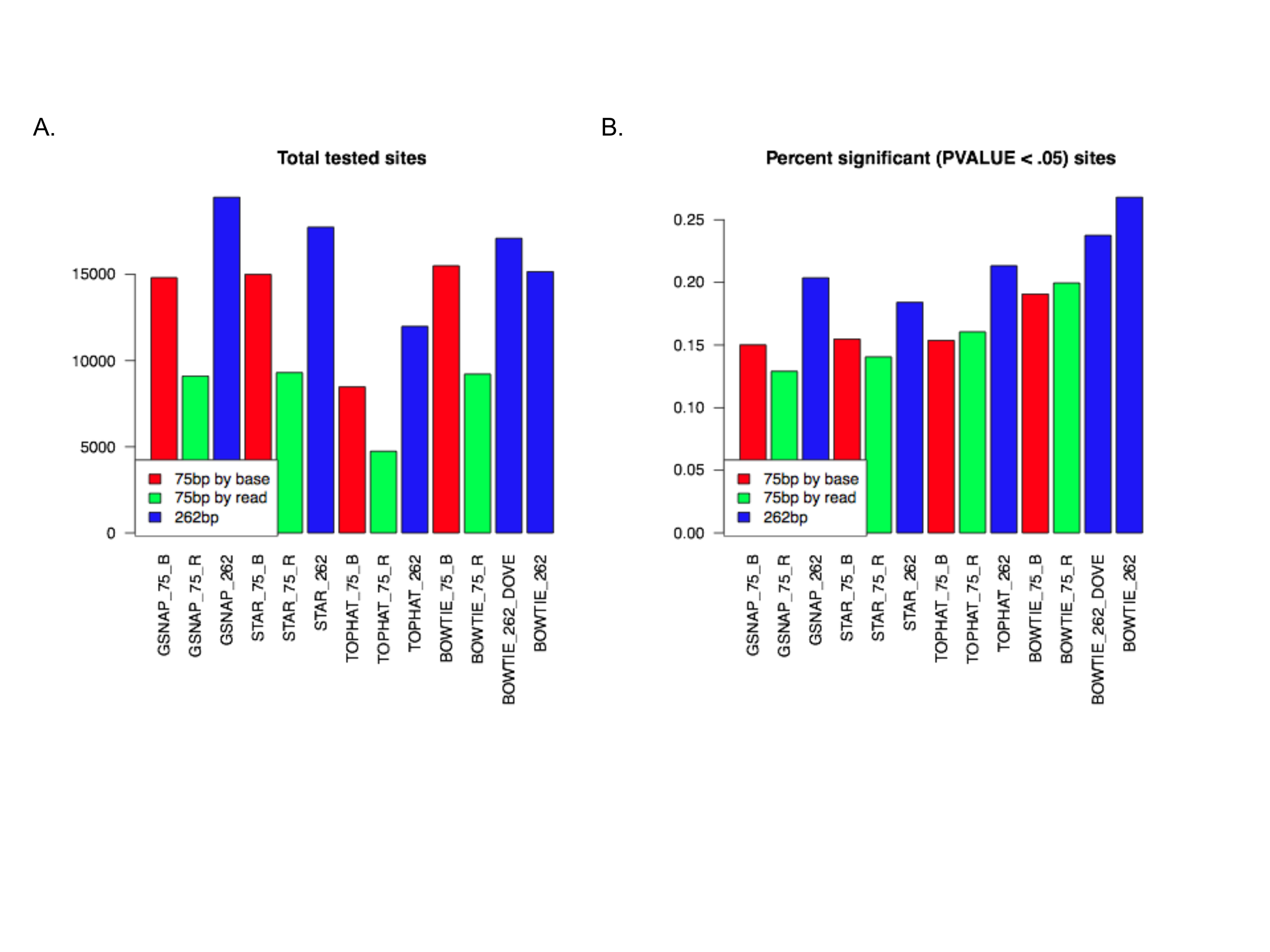}
\end{center}
\caption{
{\bf Comparison of read mapping tools and read length for ASE detection.}
A. Presents the total number of heterozygous sites tested per mapper (Bowtie, STAR, GSNAP, and Tophat) and read length (75bp and 262bp) combination. The 75bp read length data was subsampled to mirror the number of bases (red) and reads (green) in the 262bp data. The number of tested sites is very similar between STAR and GSNAP regardless of read length, while the number of tested sites for Tophat and Bowtie is significantly lower (on the order of a few thousand sites). This result is in keeping with the number of mapped reads from STAR and GSNAP versus Bowtie and Tophat. B. Percent of significant ASE sites (pvalue < .05) for each mapper and read length pair. In general, the long read samples, regardless of mapper, showed a higher percentage of significant sites. 
}
\label{fig:tested_sites}
\end{figure}

\begin{figure}[!ht]
\begin{center}
\includegraphics[width=5in]{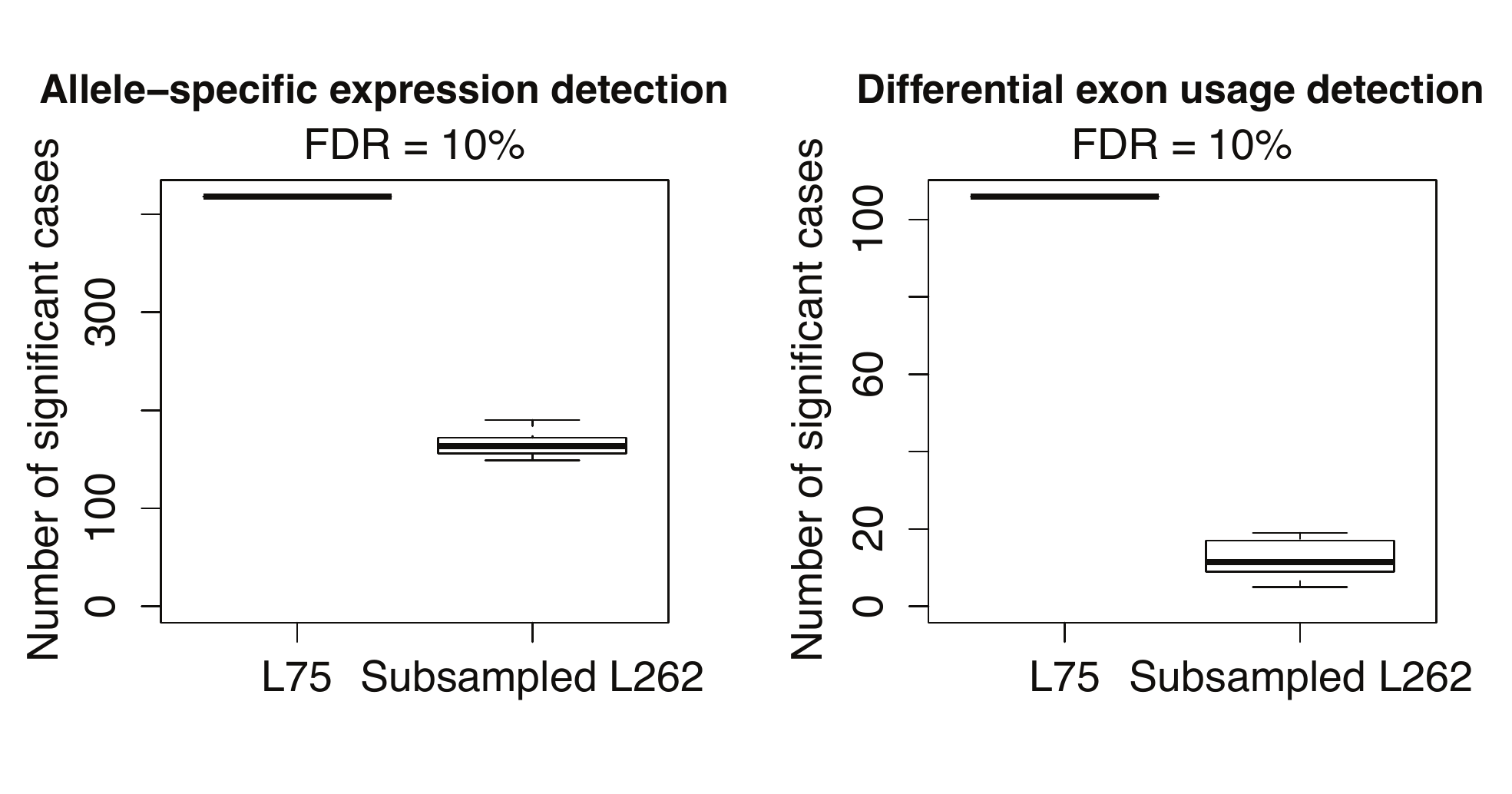}
\end{center}
\caption{
{\bf Comparison of allele-specific pattern detection with matching monetary costs.}
L262 was subsampled down to 45.45\% of the original data to match its monetary cost with that of L75. This severely reduces the amount of information contained in the subsampled L262 library, and thus results in a significantly fewer discoveries compared to L75.
}
\label{fig:ase_pattern_cost_match}
\end{figure}

\end{document}